\documentstyle[aps,epsfig,floats]{revtex}

\tighten
\renewcommand{\ol}{\overline}
\newcommand{\Pslash}{\kern 0.2 em P\kern -0.56em \raisebox{0.3ex}{/}}
\newcommand{\pslash}{\kern 0.2 em p\kern -0.4em /}
\newcommand{\kslash}{\kern 0.2 em k\kern -0.45em /}
\newcommand{\nslash}{\kern 0.2 em n\kern -0.45em /}
\newcommand{\kaslash}{\kern 0.2 em \kappa\kern -0.45em /}
\newcommand{\Sslash}{\kern 0.2 em S\kern -0.56em \raisebox{0.3ex}{/}}
\newcommand{\Mslash}{\kern 0.2 em M\kern -0.70em \raisebox{0.3ex}{/}}
\newcommand{\g}{\gamma}
\newcommand{\sig}{\sigma}
\newcommand{\eps}{\epsilon}

\newcommand{\lcvec}[3]{\left[\;#1\;,\;#2\;,\;#3\;\right]}
\newcommand{\sT}{{\scriptscriptstyle T}}
\newcommand{\nn}{\nonumber}
\newcommand{\open}{{<\kern -0.3 em{\scriptscriptstyle )}}}
\newcommand{\sumint}{\kern 0.2 em {\textstyle\sum} \kern -1.1 em \int_X}

\def\d{{\rm d}}

\begin{document}
\draft
\title{
\begin{flushright}
{\small Preprint \\
WU B 01-09}
\end{flushright}
Accessing transversity with \\
interference fragmentation functions
}

\author{Marco Radici}
\address{Dipartimento di Fisica Nucleare e Teorica, 
Universit\`{a} di Pavia, and\\
Istituto Nazionale di Fisica Nucleare, 
Sezione di Pavia, I-27100 Pavia, Italy}

\author{Rainer Jakob}
\address{Fachbereich Physik, Universit\"at Wuppertal, 
D-42097 Wuppertal, Germany}

\author{Andrea Bianconi}
\address{Dipartimento di Chimica e Fisica per i Materiali e 
per l'Ingegneria,\\ 
Universit\`{a} di Brescia, I-25133 Brescia, Italy}

\date{\today}

\maketitle

\begin{abstract}
We discuss in detail the option to access the transversity distribution 
function $h_1(x)$ by utilizing the analyzing power of interference 
fragmentation functions in two-pion production inside the same current jet. 
The transverse polarization of the fragmenting quark 
is related to the transverse component of the relative momentum of 
the hadron pair via a new azimuthal angle. 
As a specific example, we spell out thoroughly the way to extract $h_1(x)$ 
from a measured single spin asymmetry in two-pion inclusive lepton-nucleon
scattering. To estimate the sizes of observable effects we employ a spectator
model for the fragmentation functions. The resulting asymmetry of our example
is discussed as arising in different scenarios for the transversity.
\end{abstract}


\section{Introduction}\label{sec:intro}

At leading power in the hard scale $Q$, the quark content of a nucleon state is 
completely characterized by three distribution functions (DF). They describe 
the quark momentum and spin with respect to a preferred longitudinal direction 
induced by a hard scattering process. Two of them, the
momentum distribution $f_1$ and the longitudinal spin distribution $g_1$, have
been reliably extracted from experiments and accurately parametrized. Their
knowledge has deeply contributed to the studies of the quark-gluon
substructure of the nucleon. The third one, the transversity distribution
$h_1$, measures the probability difference to find the quark polarization
parallel versus antiparallel to the transverse polarization of a nucleon
target. Therefore, it correlates quarks with opposite chiralities and is
usually referred to as a ``chiral-odd'' function. Since hard scattering
processes in QCD preserve chirality at leading twist, the $h_1$ is difficult
to measure and is systematically suppressed like 
${\cal O}(1/Q)$, for example, in
inclusive deep inelastic scattering (DIS). A chiral-odd partner is needed to
filter the transversity out of the cross section.  

Historically, the so-called double spin asymmetry (DSA) in Drell-Yan 
processes with two transversely polarized protons ($p^\uparrow$) was 
suggested first~\cite{Ralston:1979ys}. However, the transversity distribution
$h_1$ for antiquarks in the proton is presumably
small~\cite{Jaffe:1997yz}. Moreover, an  upper limit for the DSA derived in a
next-to-leading order analysis by using the Soffer bounds on transversity was
found to be discouraging low~\cite{Martin:1998rz}. 

As for DIS, semi-inclusive reactions need to be considered in order to provide
the chiral-odd partner to $h_1$. In fact, in this case new functions enter the
game, the fragmentation functions (FF), which give information on the hadronic
structure complementary to the one delivered by the DF. At leading twist, the
FF describe the hadron content of quarks and, more generally, they contain
information on the hadronization process leading to the detected hadrons; as
such, they give information also on the quark content of hadrons that are not
(or even do not exist as) stable targets. The FF are also universal, but are
presently less known than the DF because a very high resolution 
and good particle identification are required in
the detection of the final state.  

Since pions are the most abundant particles detected in the calorimeter, it
would be natural to consider semi-inclusive processes where 
a single collinear pion is detected together with the final lepton 
inelastically scattered from a 
transversely polarized nucleon target. However, the $h_1$ would appear
convoluted with a chiral-odd fragmentation function only at twist three and,
therefore, suppressed like ${\cal O}(1/Q)$~\cite{Jaffe:1992ra}. It seems 
more convenient to select the more rare final state where a polarized 
$\Lambda$ decays into protons and pions~\cite{Jaffe:1993xb}. 
The analysis of the decay products reveals the $\Lambda$ polarization
and a DSA isolates at leading twist a contribution proportional to the product
of $h_1$ and a chiral-odd FF, $H_1$, which describes how a transversely
polarized quark $q^\uparrow$ fragments into a transversely polarized
$\Lambda^\uparrow$. But again, as in the case of the Drell-Yan DSA, low rates
are expected here too because of the few $\Lambda$ particles produced in a
hard reaction. Moreover, the theoretical knowledge of the mechanisms that
determine the polarization transfer $q^\uparrow \rightarrow \Lambda^\uparrow$
(i.e. of $H_1$) is not yet firmly established.  

For all these reasons, building single spin asymmetries (SSA) seems to be a
better strategy, i.e.~considering DIS or $p-p$ processes where only one
particle (the target) is transversely polarized, but selecting more
complicated final states. The most famous example is the Collins effect: the
analyzing power of the transverse polarization of the fragmenting quark is
represented by the transverse component of the momentum 
of the detected hadron with respect to the current jet axis. The typical reactions
would be, therefore, a semi-inclusive DIS, $ep^\uparrow \rightarrow e'\pi X$,
or $p^\uparrow p \rightarrow \pi X$, where the pion is detected not collinear
with the jet axis. At leading twist, a specific SSA allows for the
deconvolution of $h_1$ from the so-called 
Collins function $H_1^\perp$, the prototype of a new class of FF, the
interference FF, which are not only chiral-odd, but also {\it naive} T-odd: in
absence of two or more reaction channels with a significant relative phase,
they are forbidden by time-reversal invariance~\cite{Collins:1993kk}. 

{}From the experimental point of view, extraction of $h_1$ via the Collins
effect is quite a demanding task, because it requires the complete
determination of the transverse momentum of the detected hadron
(though first observations of a non-zero SSA have been 
reported~\cite{Airapetian:2000tv}). On the other
side, it is not sufficient to limit the theoretical analysis at leading
order. Because of the explicit dependence on an intrinsic transverse momentum,
some soft gluon divergencies (introduced by loop corrections to the tree level
result) do not cancel and must be summed up in Sudakov form factors. 
The net result is a dilution of the transverse momentum distribution of the
fragmenting quark and a final suppression of the SSA, particularly when in the
fragmentation process there is another scale very different from the hard one
$Q$, as for instance the transverse momentum of the produced hadron in the
Collins effect. The same phenomenon happens ``squared'' in $e^+e^-$ processes,
that, consequently, do not help in determining the Collins 
function~\cite{Boer:2001he}.
Moreover, modelling this interference FF by definition requires the
ability of giving a microscopic description of the relevant phase produced by
the quantum interference of different channels leading to the same detected
hadron: a very difficult task that implies a description of the structure of
the residual jet (as discussed in~\cite{Bianconi:2000cd}), or the introduction of 
dressed quark propagators~\cite{Collins:1993kk} which may be effectively modelled, 
for instance, by pion loop corrections~\cite{Bacchetta:2001di}. 

As a better alternative, the SSA with detection of two unpolarized 
leading hadrons inside the same jet was
suggested~\cite{Collins:1994ax,Collins:1994kq,Jaffe:1998hf}. In a previous
work, we have discussed the general framework for the interference FF 
arising in this case~\cite{Bianconi:2000cd}. Assuming that the residual
interactions between each leading hadron and the undetected jet is of higher
order than the one between the two hadrons themselves, the main result was
that $h_1$ gets factorized at leading twist through a novel interference FF,
$H_1^{\open}$, that relates the transverse polarization of the fragmenting
$q^\uparrow$ to the relative motion of the two detected hadrons. 
This new analyzing power, $H_1^{\open}$, filters out the $h_1$ in a very
advantageous way, because collinear factorization holds, which leads to an
exact cancellation of all collinear divergences, and makes the evolution
equations much simpler. Moreover, it is also easier to model the residual
interaction between the two hadrons only. 

In another previous work, we have presented a model
calculation for the case of the two hadrons being a $\pi$ and a $p$ with
invariant mass close to the Roper resonance~\cite{Bianconi:2000uc}. 
In the present paper we carry the calculation on
to the experimentally more relevant case of $\pi^+ \pi^-$ production with
invariant mass close to the $\rho$ resonance, and we discuss some of the
practical details for the extraction of $h_1$ from a spin asymmetry in
semi-inclusive lepton-nucleon DIS. This observable should be accessible, 
for instance, at HERMES (when the transversely polarized target will be operative) 
or even better at COMPASS (because of higher counting rates); it will also be 
a very interesting quantity for the future options in hadronic physics like 
ELFE, TESLA-N and EIC. However, as we like to emphasize, the calculation of the 
$\pi^+\pi^-$ fragmentation is process independent and could be most useful also for 
the spin physics program at RHIC, where the extraction of transversity is 
planned via a SSA in $p-p$ reactions.
  
The rest of the paper is organized as follows. In Sec.~\ref{subsec:iff} we
briefly recall the kinematics and the properties of the FF arising when a
transversely polarized quark fragments into two unpolarized leading hadrons in
the same current jet. Then, in Sec.~\ref{subsec:h1fromssa} we specialize 
the formulae
to the case of semi-inclusive lepton-nucleon DIS and detail the strategy for
building a SSA that allows for the extraction of $h_1$ at leading twist.  
In Sec.~\ref{sec:spectator} we consider the two hadrons to be two charged
pions with invariant mass around the $\rho$ resonance and we explicitly
calculate both the (process independent) interference FF and the SSA (for
semi-inclusive lepton-nucleon DIS) in the spectator model approximation. In
Sec.~\ref{sec:results} results are presented and commented. Conclusions and
outlooks are given in Sec.~\ref{sec:end}. 

\begin{figure}[h]
\begin{center} 
\psfig{file=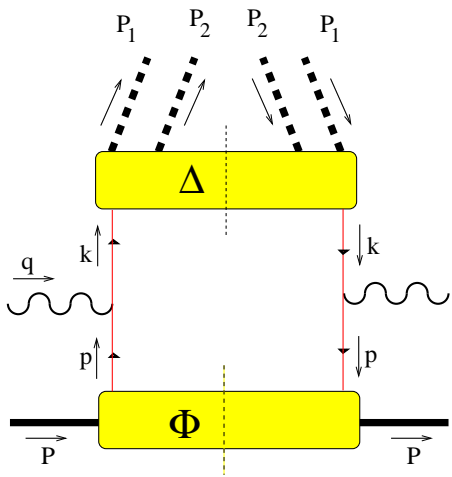, width=6cm}
\end{center} 
\caption{\label{fig:handbag} Quark diagram contributing in leading order to 
two-hadron inclusive DIS when both hadrons are in the same quark current 
jet. There is a similar diagram for anti-quarks.}
\end{figure} 

\section{Single spin asymmetry for two hadron-inclusive lepton-nucleon DIS}
\label{sec:ssa}

In this Section, we discuss the general properties of two-hadron interference 
FF when the kinematics is specialized to semi-inclusive DIS, and for this
process we work out the formula for a SSA that isolates the transversity at
leading twist. However, we emphasize that under the assumption of
factorization the soft parts of the process, i.e. the DF and the interference
FF, are universal objects and, therefore, the results can be generalized to
other hard processes, such as proton-proton scattering. 

\subsection{Interference Fragmentation Functions in semi-inclusive DIS}
\label{subsec:iff}

At leading order, the hadron tensor for two unpolarized hadron-inclusive 
lepton-nucleon DIS reads~\cite{Bianconi:2000cd}
\begin{eqnarray} 
  2M\, {\cal W}^{\mu\nu}
  &=& 
  \int \d p^-\,\d k^+\,\d^2{\vec p}_\sT^{}\,\d^2{\vec k}_\sT^{}\; 
  \delta^2\!\left({\vec p}_\sT^{}+{\vec q}_\sT^{}-{\vec k}_\sT^{}\right) 
  \mbox{Tr}\big[ 
  \; \Phi(p;P,S) \; \gamma^\mu \; \Delta(k;P_1,P_2) \; \gamma^\nu \; \big]
  \Big|_{\tiny\begin{array}{c} p^+ = x P^+ \\ k^- = P_h^-/z \end{array}}
  \nn \\ 
  & &
  {}+ \left(\begin{array}{c} 
      q\leftrightarrow -q \\ \mu \leftrightarrow \nu
    \end{array} \right) \; ,
  \label{eq:tensor}
\end{eqnarray} 
where $M$ is the target mass. The kinematics, also depicted in
Fig.~\ref{fig:handbag}, represents a nucleon with momentum $P (P^2=M^2)$ and a
virtual hard photon with momentum $q$ that hits a quark carrying a fraction
$p^+ = x P^+$ of the parent hadron momentum. We describe a 4-vector $a$ as
$\lcvec{a^-}{a^+}{{\vec a}_\sT}$ in terms of its light-cone components 
$a^\pm = (a^0\pm a^3)/\sqrt{2}$ and a transverse bidimensional vector 
${\vec a}_\sT$. Because of momentum conservation in the hard vertex, the
scattered quark has momentum $k=p+q$, and it fragments into two unpolarized
hadrons, which carry a fraction $(P_1+P_2)^-\equiv P_h^- = z k^-$ of
the ``parent quark'' momentum, and the rest of the jet.  
           
The quark-quark correlator $\Phi$ describes the nonperturbative processes that
make the parton $p$ emerge from the spin-1/2 target, and it is symbolized by
the lower shaded blob in Fig.~\ref{fig:handbag}. Using Lorentz invariance,
hermiticity and parity invariance, the partly-integrated $\Phi$ can be
parametrized at leading twist in terms of DF as  
\begin{eqnarray}
  \Phi(x,{\vec p}_\sT) 
  &\equiv & 
  \left. \int \d p^-\;\Phi(p;P,S) \right|_{p^+ = x P^+}
  \!\!\!\!=\frac{1}{2}\,\Biggl\{
  f_1\, {\nslash}_+ + 
  f_{1T}^\perp\, \epsilon_{\mu \nu \rho \sigma}\gamma^\mu 
  \frac{n_+^\nu p_\sT^\rho S_{\sT}^\sigma}{M}
  - \left(\lambda\,g_{1L}
    +\frac{({\vec p}_\sT\cdot{\vec S}_{\sT})}{M}\,g_{1T}\right)
  {\nslash}_+ \gamma_5
  \nn \\[2 mm] 
  &&
  {}- h_{1T}\,i\sigma_{\mu\nu}\gamma_5 S_{\sT}^\mu n_+^\nu
  - \left(\lambda\,h_{1L}^\perp
    +\frac{({\vec p}_\sT\cdot{\vec S}_{\sT})}{M}\,h_{1T}^\perp\right)\,
  \frac{i\sigma_{\mu\nu}\gamma_5 p_\sT^\mu n_+^\nu}{M}
  + h_1^\perp \, \frac{\sigma_{\mu\nu} p_\sT^\mu
    n_+^\nu}{M}\Biggl\} \; ,
  \label{eq:phi}
\end{eqnarray}
where the DF depend on $x, {\vec p}_\sT$ and the polarization state of the
target is fully specified by the light-cone helicity $\lambda = M S^+ / P^+$
and the transverse component ${\vec S}_\sT$ of the target spin. Similarly, the
correlator $\Delta$, symbolized by the upper shaded blob in
Fig.~\ref{fig:handbag}, represents the fragmentation of the quark into the two
detected hadrons and the rest of the current jet and can be parametrized
as~\cite{Bianconi:2000cd} 
\begin{eqnarray} 
  \Delta 
  &\equiv &
  \left.\frac{1}{4z}\int \d k^+ \; \Delta(k;P_1,P_2) 
  \right|_{k^-=P_h^-/z} \nn\\[2mm]
  &= &
  \frac{1}{4}\left\{
    D_1\,\nslash_- - 
    G_1^\perp\,
    \frac{\eps_{\mu\nu\rho\sig}\,\g^\mu\,n_-^\nu\,k_\sT^\rho\,R_\sT^\sig}
         {M_1 M_2}\,\g_5 + 
    H_1^{\open}\, \frac{\sig_{\mu\nu}\, R_\sT^\mu\, n_-^\nu}{M_1 M_2} + 
    H_1^{\perp}\, \frac{\sig_{\mu\nu}\, k_\sT^\mu\, n_-^\nu}{M_1 M_2}
  \right\} \; ,
  \label{eq:delta}
\end{eqnarray}
where $n_\pm=\lcvec{1\mp 1}{1\pm 1}{{\vec 0}_\sT}/2$ are light-cone versors 
and $R\equiv (P_1-P_2)/2$ is the relative momentum of the hadron pair. 

For convenience, we will choose a frame where, besides $\vec P_\sT = 0$, we have 
also $\vec P_{h\sT} = 0$. By defining the light-cone momentum fraction 
$\xi = P_1^-/P_h^-$, we can parametrize the final-state momenta as 
\begin{eqnarray}
  k&=&
  \lcvec{\frac{P_h^-}{z}}{z\frac{k^2+{\vec k}_\sT^2}{2P_h^-}}{{\vec k}_\sT} 
  \;,\nn\\
  P_1&=&
  \lcvec{\xi\,P_h^-}{\frac{M_1^2+{\vec R}_\sT^2}{2\,\xi\,P_h^-}}{{\vec R}_\sT}
  \;,\nn\\
  P_2&=&
  \lcvec{(1-\xi)\,P_h^-}{\frac{M_2^2+{\vec R}_\sT^2}{2\,(1-\xi)\,P_h^-}}
  {-{\vec R}_\sT} \;.
  \label{eq:vectors}
\end{eqnarray}
{}From the definition of the invariant mass of the hadron pair, i.e. $M_h^2
\equiv P_h^2 = 2 P_h^+ P_h^-$, and the on-shell condition for the two hadrons
themselves, $P_1^2=M_1^2 , P_2^2=M_2^2$, we deduce the relation 
\begin{equation}
  {\vec R}_\sT^2=\xi\,(1-\xi)\,M_h^2-(1-\xi)\,M_1^2-\xi\,M_2^2 
  \label{eq:rt2}
\end{equation}
which in turn puts a constraint on the invariant mass from the positivity
requirement ${\vec R}_\sT^2 \geq 0$:
\begin{equation}
  M_h^2 \geq \frac{M_1^2}{\xi}+\frac{M_2^2}{1-\xi} \; .
  \label{eq:mh2}
\end{equation}

\begin{figure}[h]
\begin{center} 
\psfig{file=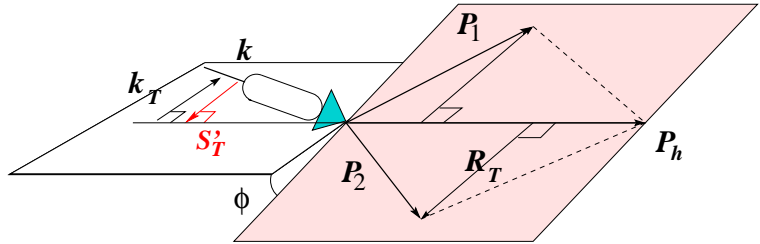, width=10cm}
\end{center} 
\caption{\label{fig:kin} The kinematics for the final state where a quark
  fragments into two leading hadrons inside the same current jet.}
\end{figure}

After having given all the details of the kinematics, we can specify the
actual dependence of the quark-quark correlator $\Delta$ and of the FF. {}From
the frame choice $\vec P_{h\sT} = 0$, the on-shell condition for both hadrons,
Eq.~(\ref{eq:rt2}), the constraint on $k^-$ and the integration over $k^+$
implied by the definition of $\Delta$ in Eq.~(\ref{eq:delta}), we deduce that
the actual number of independent components of the three 4-vectors $k,P_1,P_2$
is five (cf.~\cite{Bianconi:2000cd}). 
They can conveniently be chosen as the fraction of quark momentum
carried by the hadron pair, $z$, the subfraction in which this momentum is
further shared inside the pair, $\xi$, and the ``geometry'' of the pair in the
momentum space. Namely, the ``opening'' of the pair momenta, ${\vec R}_\sT^2$,
the relative position of the jet axis and the hadron pair axis, 
${\vec k}_\sT^2$, and the relative position of hadron pair plane and the plane
formed by the jet axis and the hadron pair axis, 
${\vec k}_\sT \cdot {\vec R}_\sT$ (see Fig.~\ref{fig:kin}). 

Both DF and FF can be deduced from suitable projections of the corresponding
quark-quark correlators. In particular, by defining
\begin{equation}
  \Delta^{[\Gamma]}
  (z,\xi,{\vec k}_\sT^2,{\vec R}_\sT^2,{\vec k}_\sT \cdot {\vec R}_\sT) 
  \equiv 
  \frac{1}{4z}\left.\int \d k^+\;\mbox{Tr}[\Gamma \, \Delta(k,P_1,P_2)] 
  \right|_{k^-=P_h^-/z} \; ,
\label{eq:proj}
\end{equation}
we can deduce
\begin{mathletters}
  \label{eq:ff}
  \begin{eqnarray}
    \Delta^{[\g^-]} &=&  
    D_1(z_h,\xi,{\vec k}_\sT^{\,2},{\vec R}_\sT^{\,2},
    {\vec k}_\sT \cdot {\vec R}_\sT)  \label{eq:d1} \\[2mm]
    \Delta^{[\g^- \g_5]}&=& 
    \frac{\eps_\sT^{ij} \,R_{\sT i}\,k_{\sT j}}{M_1\,M_2}\;
    G_1^\perp (z_h,\xi,{\vec k}_\sT^{\,2},{\vec R}_\sT^{\,2},
    {\vec k}_\sT \cdot {\vec R}_\sT)   \label{eq:g1} \\[2mm]
    \Delta^{[i\sig^{i-} \g_5]} &=& 
    {\epsilon_\sT^{ij}R_{\sT j}\over M_1+M_2}\, 
    H_1^{\open}(z_h,\xi,{\vec k}_\sT^{\,2},{\vec R}_\sT^{\,2},
    {\vec k}_\sT \cdot {\vec R}_\sT) 
    + {\epsilon_\sT^{ij}k_{\sT j}\over M_1+M_2}\,
    H_1^\perp(z_h,\xi,{\vec k}_\sT^{\,2},{\vec R}_\sT^{\,2},
    {\vec k}_\sT \cdot {\vec R}_\sT) 
    \; . \label{eq:h1}
  \end{eqnarray}
\end{mathletters}
The leading-twist projections give a nice probabilistic interpretation of FF
related to the Dirac operator $\Gamma$ used. Hence, $D_1$ is the probability
for a unpolarized quark to fragment into the unpolarized hadron pair,
$G_1^\perp$ is the probability difference for a longitudinally polarized quark
with opposite chiralities to fragment into the pair, both $H_1^\perp$ and
$H_1^{\open}$ give the same probability difference but for a transversely
polarized fragmenting quark. A different interpretation for $H_1^\perp$ and 
$H_1^{\open}$ comes only from the possible origin for a non-vanishing 
probability difference, which is induced by the direction of $k_\sT$ and 
$R_\sT$, respectively. $G_1^\perp , H_1^\perp, H_1^{\open}$ are all 
{\it naive} T-odd and $H_1^\perp, H_1^{\open}$ are further
chiral-odd. $H_1^{\open}$ represents a genuine new effect with respect to the
Collins one, because it relates the transverse polarization of the fragmenting
quark to the orbital angular motion of the transverse component of the pair
relative momentum ${\vec R}_\sT$ via the new angle $\phi$ defined by
\begin{equation}
\sin \phi = \frac{{\vec S}'_\sT \cdot {\vec P}_2 \times {\vec P}_1}
                 {|{\vec S}'_\sT| |{\vec P}_2 \times {\vec P}_1|} = 
            \frac{{\vec S}'_\sT \cdot {\vec P}_h \times {\vec R}}
                 {|{\vec S}'_\sT| |{\vec P}_h \times {\vec R}|} 
     \equiv \frac{{\vec S}'_\sT \cdot {\vec P}_h \times {\vec R}_\sT}
                 {|{\vec S}'_\sT| |{\vec P}_h \times {\vec R}_\sT|} = 
           \cos \left( \phi_{S'_\sT} - \frac{\pi}{2} - \phi_{R_\sT} \right) = 
                 \sin (\phi_{S_\sT} + \phi_{R_\sT}) \; ,
\label{eq:angle}
\end{equation}
where we have used the condition ${\vec P}_{h\sT} = 0$ and 
$\phi_{S_\sT}$ ($\phi_{S^\prime_\sT}$), $\phi_{R_\sT}$ 
are the azimuthal angles of the initial (final) quark transverse polarization 
and of ${\vec R}_\sT$ with respect to the scattering plane, respectively 
(see also Fig.~\ref{fig:kin}).

\subsection{Isolating transversity from the SSA}
\label{subsec:h1fromssa}

Usually, the analysis of experimental observables is better accomplished in
the frame where the target momentum $P$ and the momentum transfer $q$ are
collinear and with no transverse components. Using a different notation, we
have ${\vec P}_\perp = {\vec q}_\perp = 0$ and ${\vec P}_{h\perp} \neq 0$. An
appropriate transverse Lorentz boost transforms this frame to the previous one
where ${\vec P}_\sT = {\vec P}_{h\sT} = 0$ and 
${\vec q}_\sT = -{\vec P}_{h\perp}/z$~\cite{Bianconi:2000cd}. However, the 
difference between the components of vectors in each frame is suppressed like 
${\cal O}(1/Q)$. Since we are here considering expressions for the observables 
at leading twist only, this difference can be safely neglected.

\begin{figure}[h]
\begin{center} 
\psfig{file=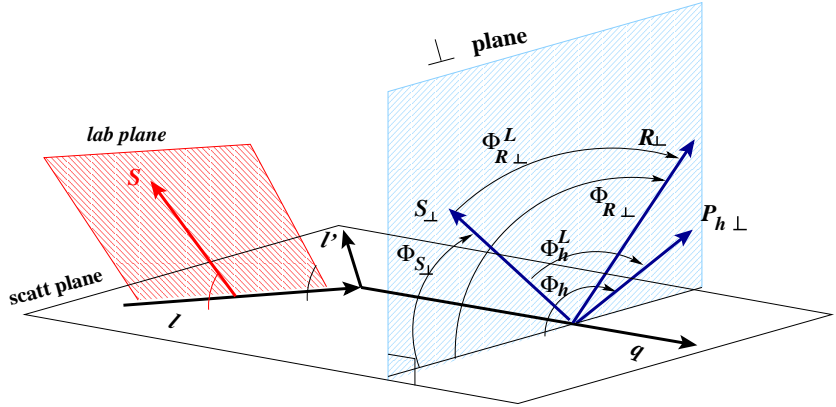, width=11cm}
\end{center} 
\caption{\label{fig:reaction} The definition of azimuthal angles, in the frame
  where $q_\perp = 0$, with respect to the scattering plane and the laboratory
  plane, whose relative oriented angle is $\phi^L = -\phi_{S_\perp}$.}
\end{figure}

By using Eq.~(\ref{eq:rt2}), the complete cross section at leading twist for
the two-hadron inclusive DIS of a unpolarized beam on a transversely polarized
target, where two unpolarized hadrons are detected in the same quark current
jet, is given by 
\begin{eqnarray}
  \lefteqn{
    \frac{\d\sigma}
         {\d\Omega\,\d x\,\d z\,\d\xi\,\d^2{\vec P}_{h\perp}\,
           \d M_h^2\,\d\phi_{R_\perp}} \, 
    =\frac{\xi (1-\xi)}{2} \, 
     \frac{\d\sigma}
          {\d\Omega\,\d x\,\d z\,\d\xi\,\d^2{\vec P}_{h\perp}\,
            \d^2{\vec R}_\perp}} 
\nn \\[2mm]
\lefteqn{
\phantom{ \frac{\d\sigma}
         {\d\Omega\,\d x\,\d z\,\d\xi\,\d^2{\vec P}_{h\perp}\,
           \d M_h^2\,\d\phi_{R_\perp}} \,}
    = \frac{\d\sigma_{OO}}
           {\d\Omega\,\d x\,\d z\,\d\xi\,\d^2{\vec P}_{h\perp}\,
             \d M_h^2\,\d\phi_{R_\perp}} \,
    + \, |{\vec S}_\perp| \, 
      \frac{\d\sigma_{OT}}
           {\d\Omega\,\d x\,\d z\,\d\xi\,\d^2{\vec P}_{h\perp}\,
             \d M_h^2\,\d\phi_{R_\perp}}}
\nn \\[2mm]
& = &
\frac{\alpha_{em} sx}{(2\pi )^3 2 Q^4} \, \Bigg\{ 
{}A(y)\;{\cal F}\left[f_1 \, D_1\right] \nn \\[2mm]
&&
{}+|{\vec R}_\perp|\;B(y)\;\sin(\phi_h+\phi_{R_\perp})\;
   {\cal F}\left[\,{\hat g}\!\cdot \!\vec p_\sT^{}\,
     \frac{h_1^{\perp} \, H_1^{\open}}{M(M_1+M_2)}\right] 
{}-|{\vec R}_\perp|\;B(y)\;\cos(\phi_h+\phi_{R_\perp})\;
   {\cal F}\left[\,{\hat h}\!\cdot \!\vec p_\sT^{}\,
     \frac{h_1^{\perp} \, H_1^{\open}}{M(M_1+M_2)}\right] \nn \\[2mm]
&& 
{}-B(y)\;\cos(2\phi_h)\;
   {\cal F}\left[\left(2\,{\hat h}\!\cdot\!\vec p_\sT^{}\,
                        \,{\hat h}\!\cdot \! \vec k_\sT^{}\,
                       -\,\vec p_\sT^{}\!\cdot \! \vec k_\sT^{}\,\right)
     \frac{h_1^{\perp} \, H_1^{\perp}}{M(M_1+M_2)}\right] \nn \\[2mm]
&&
{}-B(y)\;\sin(2\phi_h)\;
   {\cal F}\left[\left( \,{\hat h}\!\cdot\!\vec p_\sT^{}\,
                        \,{\hat g}\!\cdot \! \vec k_\sT^{}\,
                       +\,{\hat h}\!\cdot\!\vec k_\sT^{}\,
                        \,{\hat g}\!\cdot \! \vec p_\sT^{}\,\right)
     \frac{h_1^{\perp} \, H_1^{\perp}}{M(M_1+M_2)}\right] \Bigg\} \nn \\[2mm]
& + &\frac{\alpha_{em} sx}{(2\pi )^3 2 Q^4} \, |{\vec S}_\perp| \, \Bigg\{ 
A(y)\;\sin(\phi_h-\phi_{S_\perp})\;
   {\cal F}\left[\,{\hat h}\!\cdot \!\vec p_\sT^{}\,
     \frac{f_{1T}^{\perp} \, D_1}{M}\right] \, + \, 
A(y)\;\cos(\phi_h-\phi_{S_\perp})\;
   {\cal F}\left[\,{\hat g}\!\cdot \!\vec p_\sT^{}\,
     \frac{f_{1T}^{\perp} \, D_1}{M}\right]\nn\\
&& \quad 
{}+ B(y)\;\sin(\phi_h+\phi_{S_\perp})
   {\cal F}\left[\,{\hat h}\!\cdot \!\vec k_\sT^{}\,
     \frac{h_1 \, H_1^{\perp}}{M_1+M_2}\right] \, + \, 
B(y)\;\cos(\phi_h+\phi_{S_\perp})
   {\cal F}\left[\,{\hat g}\!\cdot \!\vec k_\sT^{}\,
     \frac{h_1 \, H_1^{\perp}}{M_1+M_2}\right]\nn\\
&& \quad 
{}+ |{\vec R}_\perp|\;B(y)\;\sin(\phi_{R_\perp}+\phi_{S_\perp})\;
   {\cal F}\left[\frac{h_1 \, H_1^{\open}}{M_1+M_2}\right]\nn\\
&& \quad 
{}- |{\vec R}_\perp|\;A(y)\;
   \cos(\phi_h-\phi_{S_\perp})\;\sin(\phi_h-\phi_{R_\perp})\;
   {\cal F}\left[\,{\hat h}\!\cdot \!\vec k_\sT^{}\,
                 \,{\hat h}\!\cdot \!\vec p_\sT^{}\,
     \frac{g_{1T} \, G_1^{\perp}}{MM_1M_2}\right]\nn\\
&& \quad 
{}+ |{\vec R}_\perp|\;A(y)\;
   \sin(\phi_h-\phi_{S_\perp})\;\sin(\phi_h-\phi_{R_\perp})\;
   {\cal F}\left[\,{\hat h}\!\cdot \!\vec k_\sT^{}\,
                 \,{\hat g}\!\cdot \!\vec p_\sT^{}\,
     \frac{g_{1T} \, G_1^{\perp}}{MM_1M_2}\right]\nn\\
&& \quad 
{}- |{\vec R}_\perp|\;A(y)\;
   \cos(\phi_h-\phi_{S_\perp})\;\cos (\phi_h-\phi_{R_\perp})\;
   {\cal F}\left[\,{\hat g}\!\cdot \!\vec k_\sT^{}\,
                 \,{\hat h}\!\cdot \!\vec p_\sT^{}\,
     \frac{g_{1T} \, G_1^{\perp}}{MM_1M_2}\right]\nn\\
&& \quad 
{}+ |{\vec R}_\perp|\;A(y)\;
   \sin(\phi_h-\phi_{S_\perp})\;\cos(\phi_h-\phi_{R_\perp})\;
   {\cal F}\left[\,{\hat g}\!\cdot \!\vec k_\sT^{}\,
                 \,{\hat g}\!\cdot \!\vec p_\sT^{}\,
     \frac{g_{1T} \, G_1^{\perp}}{MM_1M_2}\right]\nn\\
&& \quad 
{}+ B(y)\;\cos(3\phi_h-\phi_{S_\perp})\;
   {\cal F}\left[\,{\hat h}\!\cdot \!\vec k_\sT^{}\,
                 \,{\hat h}\!\cdot \!\vec p_\sT^{}\,
                 \,{\hat g}\!\cdot \!\vec p_\sT^{}\,
     \frac{h_{1T}^{\perp} \, H_1^{\perp}}{M^2(M_1+M_2)}\right]\nn\\
&& \quad 
{}+ B(y)\;\sin(2\phi_h)\,\cos(\phi_h-\phi_{S_\perp})\;
   {\cal F}\left[\,{\hat h}\!\cdot \!\vec k_\sT^{}\,
                 \left(\,{\hat h}\!\cdot \!\vec p_\sT^{}\,\right)^2
     \frac{h_{1T}^{\perp} \, H_1^{\perp}}{M^2(M_1+M_2)}\right]\nn\\
&& \quad 
{}- B(y)\;\cos(2\phi_h)\,\sin(\phi_h-\phi_{S_\perp})\;
   {\cal F}\left[\,{\hat h}\!\cdot \!\vec k_\sT^{}\,
                 \left(\,{\hat g}\!\cdot \!\vec p_\sT^{}\,\right)^2
     \frac{h_{1T}^{\perp} \, H_1^{\perp}}{M^2(M_1+M_2)}\right]\nn\\
&& \quad 
{}- B(y)\;\sin(3\phi_h-\phi_{S_\perp})\;
   {\cal F}\left[\,{\hat g}\!\cdot \!\vec k_\sT^{}\,
                 \,{\hat h}\!\cdot \!\vec p_\sT^{}\,
                 \,{\hat g}\!\cdot \!\vec p_\sT^{}\,
     \frac{h_{1T}^{\perp} \, H_1^{\perp}}{M^2(M_1+M_2)}\right]\nn\\
&& \quad 
{}+ B(y)\;\cos(2\phi_h)\,\cos(\phi_h-\phi_{S_\perp})\;
   {\cal F}\left[\,{\hat g}\!\cdot \!\vec k_\sT^{}\,
                 \left(\,{\hat h}\!\cdot \!\vec p_\sT^{}\,\right)^2
     \frac{h_{1T}^{\perp} \, H_1^{\perp}}{M^2(M_1+M_2)}\right]\nn\\
&& \quad 
{}+ B(y)\;\sin(2\phi_h)\,\sin(\phi_h-\phi_{S_\perp})\;
   {\cal F}\left[\,{\hat g}\!\cdot \!\vec k_\sT^{}\,
                 \left(\,{\hat g}\!\cdot \!\vec p_\sT^{}\,\right)^2
     \frac{h_{1T}^{\perp} \, H_1^{\perp}}{M^2(M_1+M_2)}\right]\nn\\
&& \quad 
{}+ |{\vec R}_\perp|\;B(y)\;
    \sin(2\phi_h+\phi_{R_\perp}-\phi_{S_\perp}) 
   {\cal F}\left[\left(({\hat h}\!\cdot\!\vec p^{}_\sT)^2
                      -({\hat g}\!\cdot\!\vec p^{}_\sT)^2
                      +2\,{\hat h}\!\cdot \!\vec p_\sT^{}\,
                        \,{\hat g}\!\cdot \!\vec p_\sT^{}\,\right)
     \frac{h_{1T}^{\perp} \, H_1^{\open}}{2M^2(M_1+M_2)}\right]
\Bigg\} \, ,
\label{eq:cross}
\end{eqnarray} 
where $\alpha_{em}$ is the fine structure constant, $s=Q^2/xy=-q^2/xy$ is the
total energy in the center-of-mass system and 
\begin{equation}
  A(y) = \left( 1-y+\frac{1}{2} y^2 \right) \quad , \quad 
  B(y) = (1-y) \quad , \quad 
  C(y) = y (2-y)
  \label{eq:abc}
\end{equation}
with the lepton invariant $y=(P \cdot q)/ (P \cdot l) \approx q^-/l^-$. The 
convolution of distribution and fragmentation functions is defined as
\begin{equation}
  {\cal F}\left[w({\vec p}_\sT^{},{\vec k}_\sT^{})\; f\, D\right]
  \equiv \;
  \sum_a e_a^2\;
  \int\d^{2}{\vec p}_\sT^{}\; \d^{2}{\vec k}_\sT^{}\;
  \delta^2 ({\vec k}_\sT^{}-{\vec p}_\sT^{}+\frac{{\vec P}_{h\perp}}{z}) \;
  w({\vec p}_\sT^{},{\vec k}_\sT^{})\;
  f^a(x,{\vec p}_\sT^{\;2})\,D^a(z_h,\xi,{\vec k}_\sT^{\,2},{\vec R}_\sT^{\,2},
  {\vec k}_\sT \cdot {\vec R}_\sT) \;,
\end{equation}
where $w({\vec p}_\sT^{},{\vec k}_\sT^{})$ is a weight function and the sum 
runs over all quark (and anti-quark) flavors, with $e_a$ the electric charges 
of the quarks. The versors appearing in the weight function $w$ are defined as 
${\hat h} = {\vec P}_{h\perp} / |{\vec P}_{h\perp}|$ and 
${\hat g}^i = \epsilon_\sT^{ij} \, {\hat h}^j$ (with 
$\epsilon_\sT^{ij} \equiv \epsilon^{-+ij}$), respectively, and they represent
the two independent directions in the $\perp$ plane perpendicular to 
${\hat z} \parallel {\vec q}/|{\vec q}|$.  All azimuthal angles 
$\phi_{S_\perp}, \phi_{R_\perp}$ and $\phi_h$ (relative to 
${\vec P}_{h\perp}$) lie in the $\perp$ plane and are measured with respect to
the scattering plane (see Fig.~\ref{fig:reaction}). Eq.~(\ref{eq:cross}) 
corresponds to the sum of Eqs. (B1) and (B4) in Ref.~\cite{Bianconi:2000cd}, 
where, however, the expressions are simpler because they rely on the 
assumption of a symmetrical cylindrical distribution of hadron pairs around 
the jet axis, in order to have fragmentation functions depending on even
powers of ${\vec k}_\sT$ only (this assumption would make all terms 
including the ${\hat g}$ versor disappear from Eq.~(\ref{eq:cross}); see 
also Ref.~\cite{Barone:2001sp} for a comparison). 

During experiments the scattering plane changes (different scales $Q$ imply
different positions of the scattered beam). Therefore, it is better to define
the laboratory frame as the plane formed by the beam and the direction of the
target polarization. All azimuthal angles are conveniently reexpressed with
respect to the laboratory frame as
\begin{eqnarray}
  \phi_{R_\perp} &= &\phi_{R_\perp}^L-\phi^L \nn \\
  \phi_{S_\perp} &= &-\phi^L \nn \\
  \phi_h &= &\phi_h^L - \phi^L \, ,
  \label{eq:azangles}
\end{eqnarray}
where the superscript $^L$ indicates the new reference frame. The oriented
angle between the scattering plane and the laboratory frame is $\phi^L$ (see
Fig.~\ref{fig:reaction}). At leading order, the azimuthal angle of 
Eq.~(\ref{eq:angle}) becomes $\phi = \phi_{R_\perp}^L- 2\phi^L$ in the new
frame.   

The new expression for the cross section is obtained by simply replacing 
Eq.~(\ref{eq:azangles}) inside the angular dependence of Eq.~(\ref{eq:cross}). 
After replacement and apart from phase space coefficients, each term of the
cross section will look like
\begin{equation}
  d\sigma^{tw} \, \propto \, t (\phi_{R_\perp}^L,\phi^L,\phi_h^L) \; 
  {\cal F} \left[ w \; \mbox{DF} \; \mbox{FF} \right] \, 
  = \, t(\phi_{R_\perp}^L,\phi^L,\phi_h^L) \; I(z,\xi,{\vec R}_\sT^{\, 2}) \, ,
  \label{eq:term}
\end{equation}
where $t$ is a trigonometric function, $w$ is the specific weight function
for each combination of distribution and fragmentation functions (DF and FF, 
respectively), and $I$ is the result of the convolution integral. It is easy 
to verify that folding the cross section by 
\begin{equation}
  \frac{1}{2\pi} \int_0^{2\pi} \d\phi^L \d\phi_{R_\perp}^L \; 
  \sin(\phi_{R_\perp}^L -2\phi^L) \; 
  \frac{\d\sigma}
       {\d\Omega\,\d x\,\d z\,\d\xi\,\d^2{\vec P}_{h\perp}\,
         \d M_h^2\,\d\phi_{R_\perp}} 
  \label{eq:fold}
\end{equation}
makes only those $d\sigma^{tw}$ terms survive where $H_1^{\open}$ shows up in
the convolution, i.e. for the following combinations
\begin{mathletters}
\label{eq:tws}
\begin{eqnarray}
  t = \cos (\phi_h^L + \phi_{R_\perp}^L -2 \phi^L) 
  &\quad , &\quad 
  w = {\hat h} \cdot {\vec p}_\sT \quad ; \label{eq:OOh} \\
  t = \sin (\phi_h^L + \phi_{R_\perp}^L -2 \phi^L) 
  &\quad , &\quad 
  w = {\hat g} \cdot {\vec p}_\sT \quad ; \label{eq:OOg} \\
  t = \sin (\phi_{R_\perp}^L - 2\phi^L) 
  &\quad , &\quad 
  w = 1 \quad ; \label{eq:OT1} \\
  t = \sin (2\phi_h^L + \phi_{R_\perp}^L - 2\phi^L) 
  &\quad , &\quad 
  w = \left(({\hat h}\cdot {\vec p}_\sT)^2 - ({\hat g}\cdot {\vec p}_\sT)^2 
    + 2 \, {\hat h}\cdot {\vec p}_\sT \, {\hat g}\cdot {\vec p}_\sT \right) 
  \label{eq:OThg} \, .
\end{eqnarray}
\end{mathletters}
Similarly, it is straightforward to proof that integrating these surviving
terms upon $\d^2{\vec P}_{h\perp}$, and performing the integrals in the
convolution ${\cal F}[w\;\mbox{DF}\; \mbox{FF}]$, makes only the
combination~(\ref{eq:OT1}) to survive presenting the transversity in a
factorized form. In fact, by integrating also upon $\d\xi$ we finally have 
\begin{eqnarray}
  \frac{<\d\sigma_{OT} >}{\d y\,\d x\,\d z\,\d M_h^2} 
  &\equiv &
  \frac{1}{2\pi} \, 
  \int_0^{2\pi} \d\phi^L\,\d\phi_{R_\perp}^L 
  \int \d^2{\vec P}_{h\perp} \;  \int \d\xi \; \sin(\phi_{R_\perp}^L -2\phi^L) \; 
  \frac{\d\sigma}{\d\Omega\,\d x\,\d z\,\d\xi\,\d M_h^2\,
    \d\phi_{R_\perp}^L\,\d^2{\vec P}_{h\perp}}  \nn \\
  &= &
  \frac{\pi \alpha_{em}^2 s x}{(2 \pi)^3 Q^4} \; 
  \frac{B(y) \, |{\vec S}_\perp |}{2(M_1+M_2)} \; \sum_a e_a^2\; 
  \int \d^2{\vec p}_\sT  \; 
  h_1^a(x,{\vec p}_\sT^{\; 2}) \nn \\
  & &\times 
  \int \d\xi \; |{\vec R}_\perp | 
  \int_0^{2 \pi} \d\phi_{R_\perp}^L 
  \int \d^2{\vec k}_\sT \; 
  H_1^{\open \, a} 
  (z,\xi,M_h^2,{\vec k}_\sT^{\; 2}, {\vec k}_\sT \cdot{\vec R}_\sT) \nn \\
  &= &
  \frac{\pi \alpha_{em}^2 s}{(2 \pi)^3 Q^4} \; 
  \frac{B(y) \, |{\vec S}_\perp |}{2(M_1+M_2)} \; 
  \sum_a e_a^2\; x \; h_1^a(x) \; 
  H_{1\, (R)}^{\open \, a} (z,M_h^2) 
  \, ,
  \label{eq:crossfact}
\end{eqnarray}
where, for sake of simplicity, the same notations are kept for DF and FF 
before and after integration, distinguished by the explicit arguments 
only; the subscript $_{(R)}$ reminds of the additional weighting factor 
$|\vec R_\perp|$. Analogously,
\begin{eqnarray}
  \frac{<\d\sigma_{OO} >}{\d y\,\d x\,\d z\,\d M_h^2} 
  &\equiv &
  \frac{1}{2\pi} \,
  \int_0^{2\pi} \d\phi^L\, \d\phi_{R_\perp}^L 
  \int \d^2{\vec P}_{h\perp} \;  \int \d\xi \;
  \frac{\d\sigma}{\d\Omega\,\d x\,\d z\,\d\xi\,\d M_h^2\,
    \d\phi_{R_\perp}^L\,\d^2{\vec P}_{h\perp}}\nn \\
  &= &
  \frac{\pi \alpha_{em}^2 s x}{(2 \pi)^3 Q^4} \, A(y) \, \sum_a e_a^2\; 
  \int \d^2{\vec p}_\sT  \; f_1^a (x,{\vec p}_\sT^{\; 2}) \nn \\
  & &\times 
  \int \d\xi \int_0^{2 \pi} \d\phi_{R_\perp}^L 
  \int \d^2{\vec k}_\sT \; 
  D_1^a (z,\xi,M_h^2,{\vec k}_\sT^{\; 2}, {\vec k}_\sT \cdot {\vec R}_\sT) 
  \nn \\
  &= &
  \frac{\pi \alpha_{em}^2 s}{(2 \pi)^3 Q^4} \, A(y) \, \sum_a e_a^2\; x \; 
  f_1^a(x) \; D_1^a (z,M_h^2) 
  \, ,
  \label{eq:crossfact2}
\end{eqnarray}
from which we can build the single spin asymmetry
\begin{eqnarray}
  A^{\sin \phi} (y,x,z,M_h^2) 
  &\equiv &
  \frac{<\d\sigma_{OT} >}{\d y\,\d x\,\d z\,\d M_h^2} \,
  \left[ \frac{<\d\sigma_{OO} >}{\d y\,\d x\,\d z\,\d M_h^2} \right]^{-1} 
  \nn \\
  &= &
  \frac{B(y)}{A(y)} \; \frac{|{\vec S}_\perp |}{2(M_1+M_2)} \; 
  \frac{\sum_a e_a^2 \; x \, h_1^a (x) \, H_{1 \, (R)}^{\open \, a} (z,M_h^2)}
  {\sum_a e_a^2 \; x \, f_1^a (x) \, D_1^a (z,M_h^2)}  
  \, .
  \label{eq:ssa}
\end{eqnarray}

\section{Spectator model for $\pi^+ \pi^-$ fragmentation}
\label{sec:spectator}

In the field theoretical description of hard processes, the FF represent the
soft processes that connect the hard quark to the detected hadrons via
fragmentation, i.e.~they are hadronic matrix elements of nonlocal operators
built from quark (and gluon) fields~\cite{Soper:1977jc}.
For a quark fragmenting into two hadrons
inside the same current jet, the appropriate quark-quark correlator (in the
light-cone gauge) reads~\cite{Collins:1994kq,Collins:1994ax}
\begin{equation}
  \Delta_{ij}(k,P_1,P_2)= \sumint \; 
  \int \frac{\d^{4\!}\zeta}{(2\pi)^4} \; e^{ik\cdot\zeta}\;
  \langle 0|\psi_i(\zeta)|P_1,P_2,X\rangle
  \langle X,P_2,P_1|\ol{\psi}_j(0)|0\rangle \, ,
  \label{eq:defDelta}
\end{equation}
where the sum runs over all the possible intermediate states containing the
hadron pair. 

\begin{figure}[h]
\begin{center}
\psfig{file=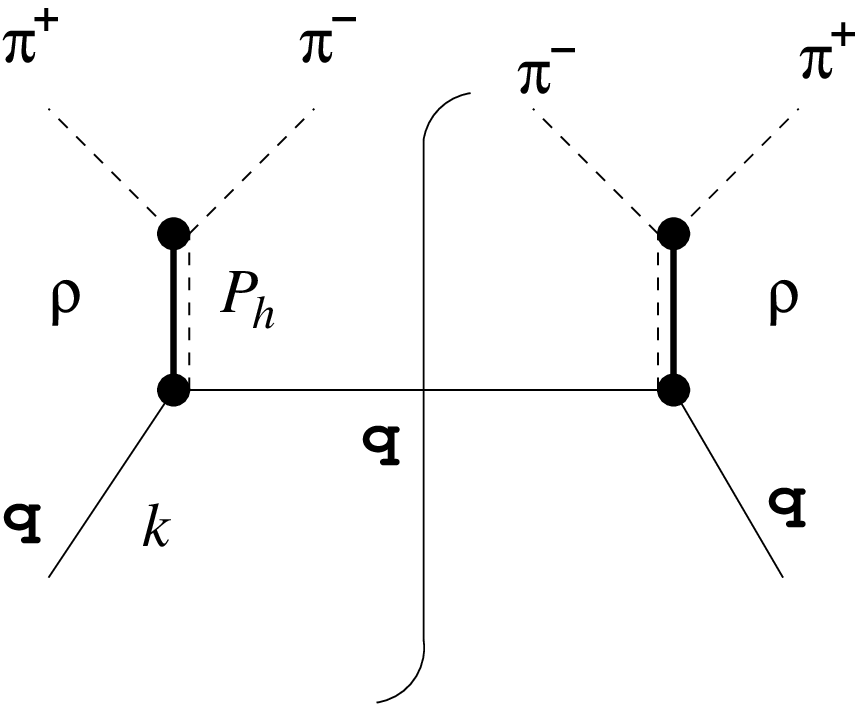, width=4.5cm}\hspace{1cm}
\psfig{file=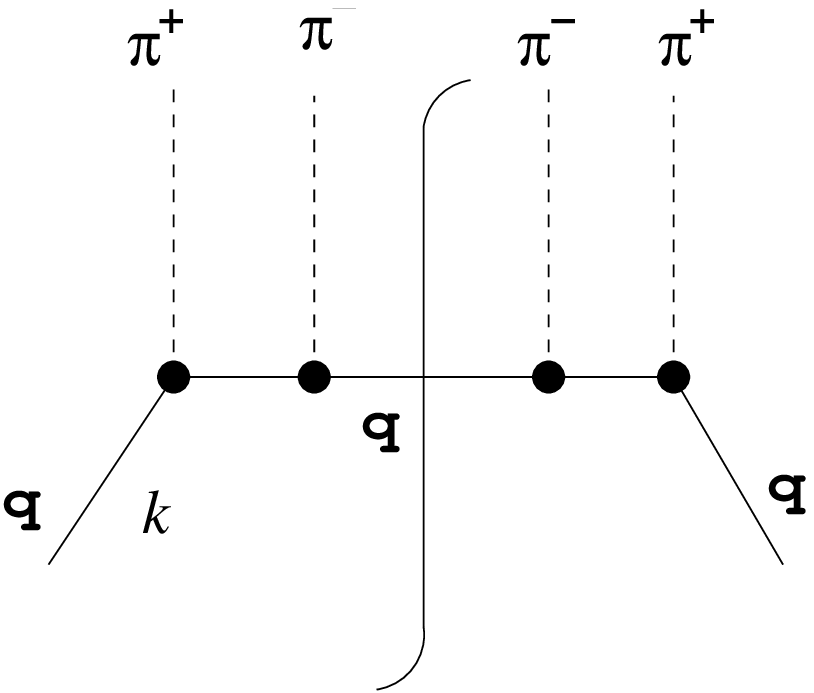, width=4.5cm}\\[0.5cm]
\psfig{file=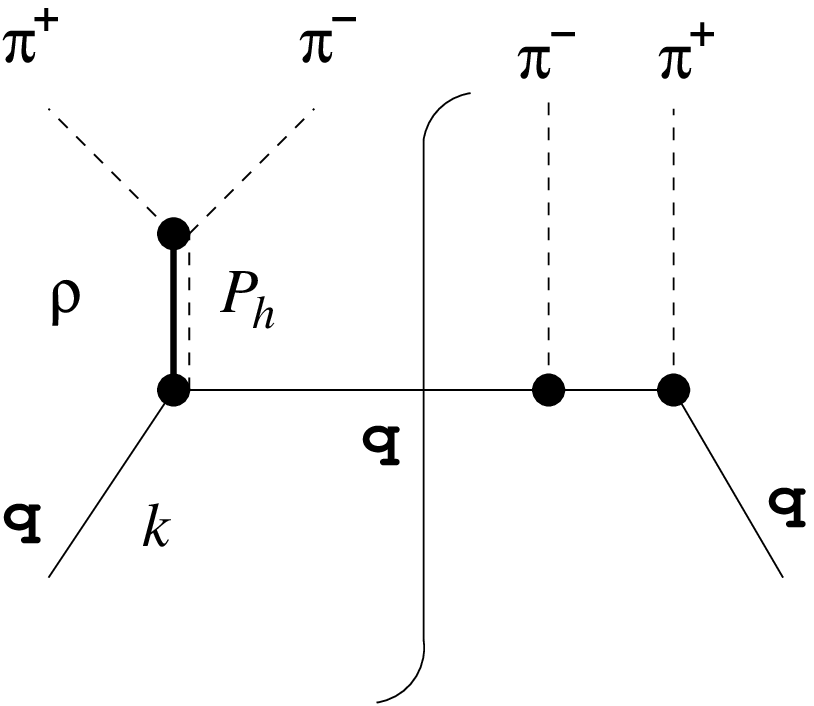, width=4.5cm}\hspace{1cm}
\psfig{file=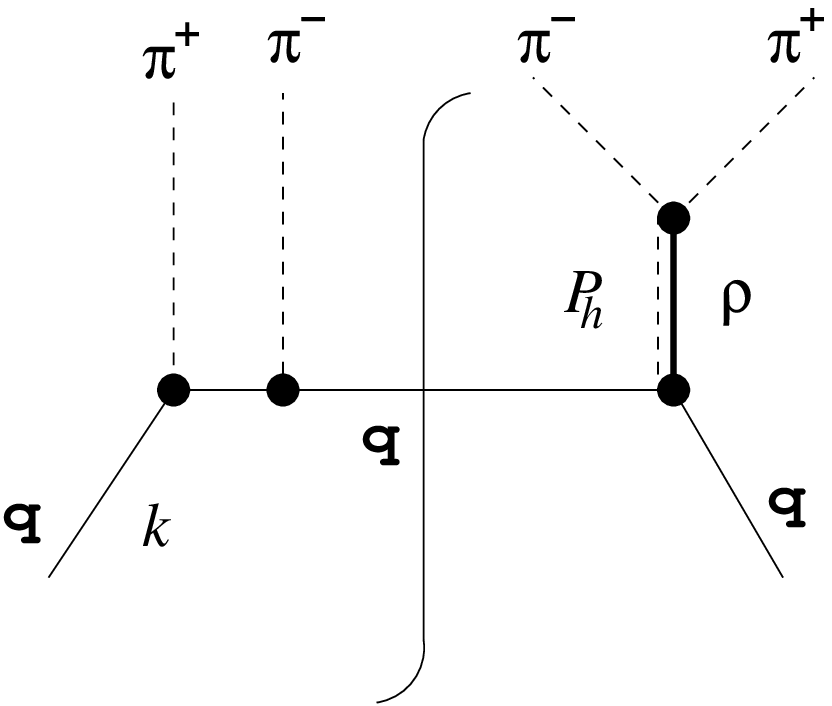, width=4.5cm}
\end{center}
\caption{\label{fig:diagspec} The diagrams considered for the quark
  fragmentation into $\pi^+ \pi^-$ at leading twist and leading order in
  $\alpha_s$ in the context of the spectator model.}
\end{figure}

The basic idea of the spectator model is to make a specific ansatz for this 
spectral decomposition by replacing the sum with an effective spectator state
with a definite mass and quantum 
numbers~\cite{Meyer:1991fr,Jakob:1997wg,Bianconi:2000uc}. By specializing 
the model to the
case of $\pi^+ \pi^-$ fragmentation with $P_1=P_{\pi^+}$ and $P_2=P_{\pi^-}$,
the spectator has the quantum numbers of an on-shell valence quark with a
constituent mass $m_q=340$ MeV. Consequently, the quark-quark correlator
(\ref{eq:defDelta}) simplifies to  
\begin{eqnarray} 
  \Delta_{ij}(k,P_{\pi^+},P_{\pi^-})
  &\approx&
  \frac{\theta\!\left((k-P_h)^+\right)}{(2\pi)^3} \; 
  \delta\left((k-P_h)^2-m_q^2\right)\;
  \langle 0|\psi_i(0)|P_{\pi^+},P_{\pi^-},q\rangle
  \langle q,P_{\pi^-},P_{\pi^+}|
  \ol{\psi}_j(0)|0\rangle  \nn\\
  &\equiv&
  \widetilde\Delta_{ij}(k,P_{\pi^+},P_{\pi^-})\;
  \delta(\tau_h-\sig_h+M_h^2-m_q^2) \;,
  \label{eq:specDelta}
\end{eqnarray}
where $\tau_h = k^2$ and $\sigma_h = 2k\cdot P_h$. When inserting 
Eq.~(\ref{eq:specDelta}) into Eq.~(\ref{eq:proj}), the projections drastically
simplify to 
\begin{equation} 
  \Delta^{[\Gamma]}(z_h,\xi,{\vec k}_\sT^{\,2},\vec R_\sT^{\, 2},
  {\vec k}_\sT \cdot {\vec R}_\sT)=
  \left.\frac{\mbox{Tr}[\Gamma \, \widetilde\Delta]}{8(1-z)P_h^-}
  \right|_{\tau_h=\tau_h(z,{\vec k}_\sT^{\,2})} \; ,
  \label{eq:projspect}
\end{equation} 
with
\begin{equation}
  \tau_h(z,{\vec k}_\sT^{\,2})=\frac{z}{1-z}{\vec k}_\sT^{\,2}
  +\frac{m_q^2}{1-z}+\frac{M_h^2}{z} \;.
  \label{eq:tauspect}
\end{equation} 

We will consider the $\pi^+ \pi^-$ system with an invariant mass $M_h$ close
to the $\rho$ resonance, specifically 
$m_\rho - \Gamma_\rho/2 \leq M_h \leq m_\rho + \Gamma_\rho/2$, where 
$\Gamma_\rho$ is the width of the $\rho$ resonance. Hence, the
most appropriate and simplest diagrams that can replace the quark decay of
Fig.~\ref{fig:handbag} at leading twist, and leading order in $\alpha_s$, are
represented in Fig.~\ref{fig:diagspec}: the $\pi^+ \pi^-$ can be produced from
the $\rho$ decay or directly via a quark exchange in the $t$-channel (the
background diagram); the quantum interference of the two processes generates
the {\it naive} T-odd FF described in Sec.~\ref{subsec:iff}. A suitable 
selection
of ``Feynman'' rules for the vertices and propagators of the diagrams in
Fig.~\ref{fig:diagspec} allows for the analytic calculation of the matrix
elements defining $\widetilde \Delta$ in Eq.~(\ref{eq:specDelta}) and,
consequently, of the projections $\Delta^{[\Gamma]}$ defining the FF. 

\subsection{Propagators}
\label{sec:propagators}

The propagators involved in the diagrams of Fig.~\ref{fig:diagspec} are:

\begin{itemize} 
\item quark with momentum $\kappa$
      \begin{center}
      \psfig{file=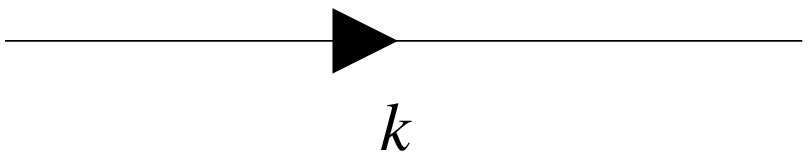, width=3cm} 
      \hspace{3mm} 
      \begin{minipage}{60mm}
      \begin{eqnarray*}
       \left(\frac{i}{\kaslash -m_q}\right)_{ij}
       \end{eqnarray*}
       \vspace{1mm}
       \end{minipage}
       \end{center} \par \noindent
       The propagator occurs with $\kappa^2 = \tau_h \equiv k^2$ or 
       $\kappa^2 = (k-P_{\pi^+})^2$. In both cases, the off-shell condition 
       $k^2 \neq m_q^2$ is guaranteed by Eq.~(\ref{eq:tauspect}).
       
\item $\rho$ with momentum $P_h$
      \begin{center}
      \psfig{file=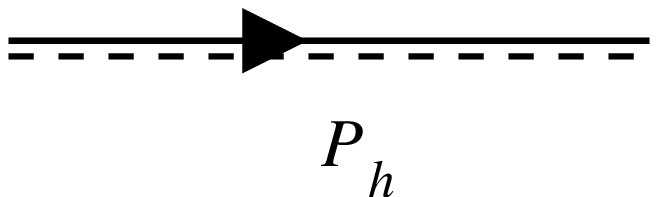, width=3cm} 
      \hspace{3mm} 
      \begin{minipage}{60mm}
      \begin{eqnarray*}
       & &\frac{i}{P_h^2-m_\rho^2+im_\rho\Gamma_\rho}
          \left(-g^{\mu\nu}+\frac{P_h^\mu\,P_h^\nu}{P_h^2}\right)
      \end{eqnarray*}
       \vspace{1mm}
       \end{minipage}
       \end{center} \par \noindent
      where $\Gamma_{\rho} = \displaystyle{\frac{f^2_{\rho \pi \pi}}{4\pi} 
      \frac{m_{\rho}}{12} \left( 1 - \frac{4 m_{\pi}^2}{m_{\rho}^2} 
      \right)^{\frac{3}{2}}}$~\cite{Ioffe:1984ep}.
\end{itemize}

\subsection{Vertices}
\label{sec:vertices}

In analogy with previous works on spectator 
models~\cite{Jakob:1997wg,Bianconi:2000uc}, we choose the vertex form
factors to depend on one invariant only, generally denoted $\kappa^2$, that
represents the virtuality of the external entering quark line. Therefore, we
can have $\kappa^2 = \tau_h \equiv k^2$ or $\kappa^2 = (k-P_{\pi^+})^2$. The
power laws are such that the asymptotic behaviour is in agreement with the
expectations based on dimensional counting rules. Finally, the normalization
coefficients have dimensions such that 
$\int \d^2{\vec k}_\sT \int \d^2{\vec R}_\sT \, 
D_1(z,\xi,{\vec k}_\sT^2, {\vec R}_\sT^2,{\vec k}_\sT \cdot {\vec R}_\sT)$ 
is a pure number to be interpreted as the probability for the hadron pair to
carry a $z$ fraction of the valence quark momentum and to share it in $\xi$
and $1-\xi$ parts. 

\begin{itemize}
\item $\rho \pi \pi$ vertex
      \begin{center}
      \psfig{file=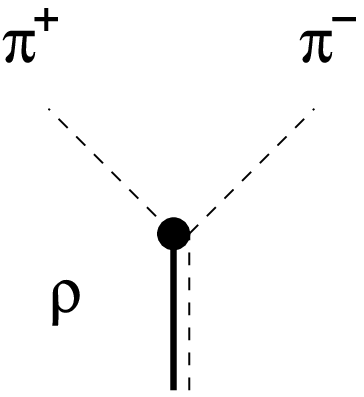, width=1.5cm} 
      \hspace{3mm} 
      \begin{minipage}{60mm}
      \begin{eqnarray*}
       & &\Upsilon^{\rho\pi\pi, \mu} = f_{\rho\pi\pi} R^\mu 
      \end{eqnarray*}
      \vspace{1mm}
      \end{minipage}
      \end{center} \par \noindent 
     where $\displaystyle{\frac{f_{\rho\pi\pi}^2}{4\pi}}=2.84\pm 0.50$~\cite{Ioffe:1984ep}.
     
\item $q \rho q$ vertex
      \begin{center}
      \psfig{file=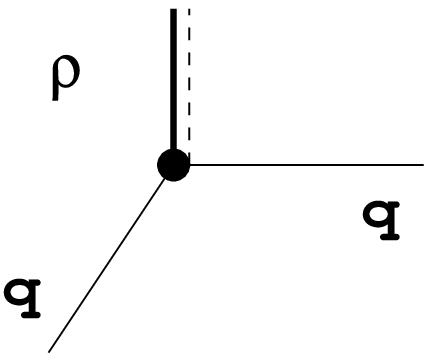, width=1.5cm} 
      \hspace{3mm} 
      \begin{minipage}{60mm}
      \begin{eqnarray*}
       \Upsilon^{q\rho q,\mu}_{ij}&=&\frac{f_{q\rho q}
       (\kappa^2)}{\sqrt{2}} \; [\g^\mu]_{ij} \nn \\
       &= &\frac{N_{q\rho}}{\sqrt{2}} \; 
       \frac{1}{|\kappa^2-\Lambda_{\rho}^2|^{\alpha}} \; 
       [\g^\mu]_{ij}
      \end{eqnarray*}
      \vspace{1mm}
      \end{minipage}
      \end{center} \par \noindent 
      where $\Lambda_\rho$ 
      excludes large virtualities of the quark. The power 
      $\alpha$ is determined consistently with the quark counting rule that 
      determines the asymptotic behaviour of the FF at large 
      $z$~\cite{Ioffe:1984ep}, i.e.
      \begin{equation}
      (1-z)^{2\alpha -1} = (1-z)^{-3+2r+2|\lambda|} \, ,
      \label{eq:alpha}
      \end{equation}
      where $r$ is the number of constituent quarks in the considered 
      hadron, and $\lambda$ is the difference between the quark and the 
      hadron helicities. 
      Thus, here we have $\alpha = 3/2$. The normalization
      $N_{q\rho}$ is such that the sum rule
      \begin{equation}
        \int_0^1 \d z \; z\,D_1(z) \leq 1 
      \label{eq:sumrule}
      \end{equation}
      is satisfied. 
      In fact, in the infinite momentum frame 
      the integral in Eq.~(\ref{eq:sumrule}) represents the 
      total fraction $z$ of the quark energy taken by all hadron pairs
      of the type under consideration. Since in this frame low-energy
      mass effects can be neglected, we estimate that charged pion pairs with
      an invariant mass inside the $\rho$ resonance width represent $\sim
      50\%$ of the total pions detected in the calorimeter, which in turn can
      be considered $\sim 80\%$ of all particles detected. 
      Neglecting mass effects, we may assume that the fraction of quark 
      energy taken by charged pions, relative to the energy taken by other 
      hadrons, follows their relative numbers. Therefore, we chose
      two values, $N_{q\rho} = 0.9$ GeV$^3$ and $1.6$ GeV$^3$, which
      correspond to rather extreme scenarios where the 
      integral~Eq.~(\ref{eq:sumrule}) amounts to 0.14 and 0.48, respectively. 
      
\item $q \pi q$ vertex
      \begin{center}
      \psfig{file=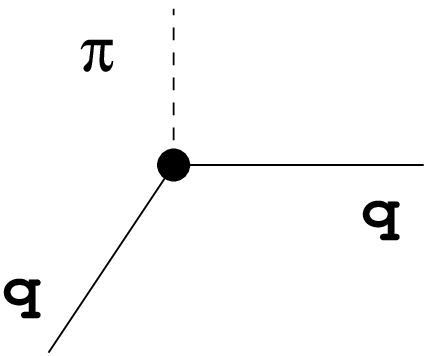, width=1.5cm} 
      \hspace{3mm} 
      \begin{minipage}{60mm}
      \begin{eqnarray*}
       \Upsilon^{q\pi q}_{ij}&=&\frac{f_{q\pi q} (\kappa^2)}{\sqrt{2}} \;
       [\g_5]_{ij} \nonumber \\
       &=
       &\frac{N_{q\pi}}{\sqrt{2}} \, 
       \frac{1}{|\kappa^2-\Lambda_{\pi}^2|^\alpha} 
       \;[\g_5]_{ij}
       \end{eqnarray*}
       \vspace{1mm}
       \end{minipage}
       \end{center} \par \noindent 
       where $\Lambda_\pi$ excludes large virtualities of the quark, as
       well. {}From quark counting rules, still $\alpha = 3/2$. The
       normalization $N_{q\pi}$ can be deduced from $N_{q\rho}$ by
       generalizing the Goldberger-Treiman relation to the $\rho$-quark
       coupling~\cite{Glozman:1998fs}: 
       \begin{eqnarray}
       \frac{g_{\pi qq}^2}{4\pi} &= &\left( \frac{g_q^A}{g_N^A} \right)^2 
          \left( \frac{m_q}{m_N} \right)^2 \frac{g_{\pi NN}^2}{4\pi} = 
          \left( \frac{3}{5} \right)^2 \left( \frac{340}{939} \right)^2 14.2 
          = 0.67 
          \nn \\
       \frac{(g_{\rho qq}^V+g_{\rho qq}^T)^2}{4\pi} &= &
       \left( \frac{g_q^A}{g_N^A} 
         \right)^2 \left( \frac{m_q}{m_N} \right)^2 
         \frac{(g_{\rho NN}^V+g_{\rho NN}^T)^2}{4\pi} = 
         \left( \frac{3}{5} \right)^2 
         \left( \frac{340}{939} \right)^2 27.755 = 1.31 \, , \label{eq:g-t}
       \end{eqnarray}
       where $g^A_N, m_N$ are the nucleon axial coupling constants and mass,
       respectively, as well as $g^A_q, m_q$ the quark ones. The $\pi NN$
       coupling is $g_{\pi NN}^2 /4\pi = 14.2$; the vector $\rho NN$ coupling
       is $(g_{\rho NN}^V)^2 /4\pi = 0.55$ and its ratio to the tensor
       coupling is 
       $g_{\rho NN}^T / g_{\rho NN}^V=6.105$~\cite{Ericson:1988gk}. 
       {}From the above relations, we deduce
       \begin{equation}
       \frac{g_{\pi qq}}{(g_{\rho qq}^V+g_{\rho qq}^T)} \equiv 
         \frac{N_{q\pi}}{N_{q\rho}} = 0.715 \, .
       \label{eq:qpi-coup}
       \end{equation}
\end{itemize}

As a final comment, we have explicitly checked that with the above rules the
background diagram leads to a cross section that qualitatively shows the same
$s$ dependence of experimental data for $\pi \pi$ production in the relative
$L=0$ channel when $s$ is inside the $\rho$ resonance width, in any case below
the first dip corresponding to the resonance 
$f_0 (980)$~\cite{Pennington:1999fa}. If 
we reasonably assume that the resonant diagram exhausts almost all of 
the $\pi \pi$
production in the relative $L=1$ channel and we also assume that in the given
energy interval the $L=0,1$ channels approximate the whole strength for $\pi
\pi$ production, we can safely state that the diagrams of
Fig.~\ref{fig:diagspec} give a satisfactory reproduction of the $\pi \pi$
cross section, with invariant mass in the given interval, without invoking any
scalar $\sigma$ resonance (cf.~\cite{Collins:1994ax,Jaffe:1998hf}).

\subsection{Interference FF}
\label{sec:speciff}

With the above rules applied to the diagrams of Fig.~\ref{fig:diagspec}, we
can calculate all the matrix elements of Eq.~(\ref{eq:specDelta}) and,
consequently, all the projections (\ref{eq:projspect}) leading to the FF. The
{\it naive} T-odd $G_1^\perp, H_1^\perp, H_1^{\open}$ receive contributions
from the interference diagrams only. In particular, they result proportional
to the imaginary part of the $\rho$ propagator ($\sim m_\rho \Gamma_\rho$),
while the real part ($\sim M_h^2 - m_\rho^2$) contributes to $D_1$. Therefore,
contrary to the findings of Ref.~\cite{Jaffe:1998hf}, 
a complex amplitude with a resonant behaviour is
needed here to produce nonvanishing interference FF. For a $u$ quark 
fragmenting into $\pi^+ \pi^-$, we have at leading twist 
\begin{eqnarray}
  \lefteqn{
    D_1^{u\rightarrow \pi^+ \pi^-} 
    (z,\xi,M_h^2,k_\sT^2,{\vec k}_\sT \cdot {\vec R}_\sT) \;=\; 
    \frac{N_{q\rho}^2 \, f_{\rho\pi\pi}^2 \, z^2 \, (1-z)^2}
    {4(2\pi)^3 \, [(M_h^2-m_\rho^2)^2+m_\rho^2\,\Gamma_\rho^2] \, 
      a^2 \, \vert a+b\vert^3} \nn}\\
  & & \qquad\qquad\times \Bigg\{ 
  \frac{c}{4} [ \, c-za(2\xi -1) ] 
  + z^2 (1-z) \left( \frac{M_h^2}{4}-m_\pi^2 \right) [ \, a-(1-z)M_h^2 \, ] 
  \Bigg\} \nn \\[3mm]
  & &
  {}+ \frac{N_{q\pi}^4 \, z^7 \, (1-z)^7}
  {8 (2\pi )^3 \, a^2 \, d^2 \, 
    \vert d+{\tilde b} \vert^3 \, 
    \vert a+{\tilde b}\vert^3} 
  \nn \\[2mm]
  & & \qquad\times \Bigg\{ 
  -az \, [ z\xi (1-z) + (1-\xi) d ] -z(1-z) m_\pi^2 \, 
  \frac{a-c-z(1-z)M_h^2}{2} \, + \, d \, \frac{a-c+z(1-z)M_h^2}{2} 
  \Bigg\}\nn\\[3mm]
  & & 
  {} + \frac{\sqrt{2} \, (M_h^2-m_\rho^2) \, z^{\frac{9}{2}} \, 
    (1-z)^{\frac{9}{2}} \, N_{q\pi}^2 \, N_{q\rho} \, f_{\rho \pi \pi}}
  {8 (2\pi)^3 [(M_h^2-m_\rho^2)^2+m_\rho^2\,\Gamma_\rho^2] \, a^2 \, d \, 
    \vert a+b \vert^{\frac{3}{2}} \, 
    \vert a+{\tilde b} \vert^{\frac{3}{2}} \, 
    \vert d+{\tilde b} \vert^{\frac{3}{2}}} \nn\\[2mm]
  & & \qquad\times \Bigg\{ 
  az(1-z) \left( 2m_\pi^2 - \frac{M_h^2}{2}\right)  
  + \frac{a(1-2z\xi) +c +z(1-z)M_h^2}{4} \, [ \, d+z(1-z) (M_h^2-5m_\pi^2) ] 
  \nn \\[2mm]
  & & \qquad\qquad
  {}+ \, \frac{a \, [2z(1-\xi)-1]+c-z(1-z)M_h^2}{4} \, 
  [\, 3d -a +z(1-z)m_\pi^2 \, ] 
  \Bigg\} 
  \label{eq:d1spec}\\[5mm]
  \lefteqn{
    H_1^{\open \, u\rightarrow \pi^+ \pi^-} 
    (z,\xi,M_h^2,k_\sT^2,{\vec k}_\sT \cdot {\vec R}_\sT) =} \nn\\[2mm]
  & &
  {}-\, \frac{m_\rho \, \Gamma_\rho \, m_\pi \, m_q \, z^{\frac{13}{2}} \, 
    (1-z)^{\frac{11}{2}} \, N_{q\pi}^2 \, N_{q\rho} \, f_{\rho \pi \pi}}
  {2\sqrt{2} (2\pi)^3 [(M_h^2-m_\rho^2)^2+m_\rho^2\Gamma_\rho^2] \, 
    a \, d \, \vert a+b\vert^{\frac{3}{2}} \, 
    \vert a+{\tilde b}\vert^{\frac{3}{2}} \, 
    \vert d + z(1-z)(m_q^2-\Lambda_\pi^2)\vert^{\frac{3}{2}}} 
  \label{eq:h1spec} \\[5mm]
  \lefteqn{
    H_1^{\perp \, u\rightarrow \pi^+ \pi^-} 
  (z,\xi,M_h^2,k_\sT^2,{\vec k}_\sT \cdot {\vec R}_\sT) = 0 }
  \label{eq:h10} \\[5mm]
  & &
  G_1^{\perp \, u\rightarrow \pi^+ \pi^-} 
  (z,\xi,M_h^2,k_\sT^2,{\vec k}_\sT \cdot {\vec R}_\sT) = 
  - \frac{m_\pi}{2m_q} \, H_1^{\open \, u\rightarrow \pi^+ \pi^-} 
  (z,\xi,M_h^2,k_\sT^2,{\vec k}_\sT \cdot {\vec R}_\sT) 
  \, , 
  \label{eq:g1perp}
\end{eqnarray}
where 
\begin{eqnarray}
  a & = & z^2(k_\sT^2+m_q^2)+(1-z) M_h^2 
  \quad , \quad 
  b= z(1-z)(m_q^2-\Lambda_\rho^2) 
  \quad , \quad 
  {\tilde b} = z(1-z)(m_q^2 - \Lambda_\pi^2) \nn \\
  c & = &(2\xi -1) [z^2(k_\sT^2+m_q^2) -(1-z)^2M_h^2]
  -4z(1-z) {\vec k}_\sT \cdot {\vec R}_\sT \nn \\
  d & = & z^2(1-\xi) (k_\sT^2+m_q^2) +\xi (1-z)^2 M_h^2 
  +z(1-z)(m_\pi^2 + 2{\vec k}_\sT \cdot {\vec R}_\sT) 
  \, .
  \label{eq:coeffs}
\end{eqnarray}

The simplifications induced by the spectator model reduce the number of
independent FF, Eq.~(\ref{eq:g1perp}), and make $H_1^\perp$ vanish,
i.e.~the analogue of the Collins effect in this context turns out to be a 
higher-order effect. The structure induced by the model is simply not rich
enough to produce a non-vanishing $H_1^\perp$. Moreover, the FF do not depend
on the flavor of the fragmenting valence quark, provided that the charges of
the final detected pions are selected according to the diagrams of
Fig.~\ref{fig:diagspec}. Hence, the FF are the same for 
$u\rightarrow \pi^+ \pi^-$ and for $d\rightarrow \pi^- \pi^+$, 
where the final state differs only by the interchange of the two pions,
i.e. by leaving everything unaltered but 
${\vec R}_\sT \rightarrow -{\vec R}_\sT$ and $\xi \rightarrow (1-\xi)$:
\begin{eqnarray}
  D_1^{\,u\rightarrow \pi^+\pi^-}
  (z,\xi,M_h^2,k_\sT^2,{\vec k}_\sT \cdot {\vec R}_\sT)
  & = &
  D_1^{\,d\rightarrow \pi^- \pi^+}
  (z,\xi,M_h^2,k_\sT^2,{\vec k}_\sT \cdot {\vec R}_\sT)
  \nn\\
  & = &
  D_1^{\,d\rightarrow \pi^+ \pi^-}
  (z,(1-\xi),M_h^2,k_\sT^2,{\vec k}_\sT \cdot(-{\vec R}_\sT))
  \nn\\[2mm]
  H_1^{\open\,u\rightarrow \pi^+ \pi^-}
  (z,\xi,M_h^2,k_\sT^2,{\vec k}_\sT \cdot {\vec R}_\sT)
  & = &
  H_1^{\open\,d\rightarrow \pi^- \pi^+}
  (z,\xi,M_h^2,k_\sT^2,{\vec k}_\sT \cdot {\vec R}_\sT)
  \nn\\
  & = &
  H_1^{\open \, d\rightarrow \pi^+ \pi^-}
  (z,(1-\xi),M_h^2,k_\sT^2,{\vec k}_\sT \cdot (-{\vec R}_\sT))
  \, .
  \label{eq:flavorsym}
\end{eqnarray}
When integrating the FF over $\d^2{\vec k}_\sT$ and $\d \xi$, the dependence on
the direction of ${\vec R}_\sT$ is lost
\begin{eqnarray}
  D_1^{\,u\rightarrow \pi^+ \pi^-}(z,M_h^2)
  & \equiv &
  \int_0^1\d\xi
  \int \d^2{\vec k}_\sT \;
  D_1^{\,u\rightarrow \pi^+ \pi^-}
  (z,\xi,M_h^2,k_\sT^2,{\vec k}_\sT \cdot {\vec R}_\sT)
  \nn\\
  & = &
  \int_0^1\d\xi
  \int \d^2{\vec k}_\sT \;
  D_1^{\,d\rightarrow \pi^+ \pi^-}
  (z,(1-\xi),M_h^2,k_\sT^2,{\vec k}_\sT \cdot (- {\vec R}_\sT))
  \nn\\
  & = &
  \int_0^1\d\xi
  \int \d^2{\vec k}_\sT \;
  D_1^{\,d\rightarrow \pi^+ \pi^-}
  (z,\xi,M_h^2,k_\sT^2,{\vec k}_\sT \cdot (- {\vec R}_\sT))
  \nn\\
  & \equiv &
  D_1^{\,d\rightarrow \pi^+ \pi^-}(z,M_h^2)
  \, ,
\end{eqnarray}
and similarly for $H_1^{\open}$. Therefore, we can conclude
that the integrated FF do not depend
in general on the flavor of the fragmenting quark.

Consequently, the SSA of Eq.~(\ref{eq:ssa}) simplifies to
\begin{equation}
  A^{\sin \phi} (y,x,z,M_h^2) = 
  \frac{B(y)}{A(y)} \; 
  \frac{|{\vec S}_\perp |}{4m_\pi} \; 
  \frac{\left[ \displaystyle{
        \frac{8}{9} x \, h_1^u (x) + \frac{1}{9} x \, h_1^d (x) }\right] \; 
    H_{1 \, (R)}^{\open \, u} (z,M_h^2)}
  {\left[ \displaystyle{
        \frac{8}{9} x \, f_1^u (x) + \frac{1}{9} x \, f_1^d (x) }\right] \;  
    D_1^u (z,M_h^2)}  
  \, .
  \label{eq:ssaspec}
\end{equation}
In the following, we will discuss the SSA without the inessential 
$|{\vec S}_\perp| B(y)/A(y)$ factor and after integrating away the $z$ 
dependence and, in turn, the $x$ or $M_h$ dependence according to
\begin{eqnarray}
  A^{\sin \phi}_x (x) 
  &\equiv &
  \frac{1}{4m_\pi} \; 
  \frac{\displaystyle{ \left[ 
        \frac{8}{9} x \, h_1^u (x) + \frac{1}{9} x \, h_1^d (x) \right] \; 
      \int \d z \,\d M_h^2 }\; 
    H_{1 \, (R)}^{\open \, u} (z,M_h^2)}
  {\displaystyle{\left[ 
        \frac{8}{9} x \, f_1^u (x) + \frac{1}{9} x \, f_1^d (x) \right] \; 
      \int \d z \,\d M_h^2 }\; 
    D_1^u (z,M_h^2)}  
  \label{eq:ssa-x} \\[2mm]
  A^{\sin \phi}_{M_h} (M_h) 
  &\equiv &
  \frac{1}{4m_\pi} \; 
  \frac{\displaystyle{\int \d x \, 
      \left[ \frac{8}{9} x \, h_1^u (x) + \frac{1}{9} x \, h_1^d (x) \right]\;
      \int \d z }\; H_{1 \, (R)}^{\open \, u} (z,M_h^2)}
  {\displaystyle{\int \d x \, 
      \left[ \frac{8}{9} x \, f_1^u (x) + \frac{1}{9} x \, f_1^d (x) \right]\;
      \int \d z }\; D_1^u (z,M_h^2)}  
  \label{eq:ssa-mh} 
  \, .
\end{eqnarray}

\vspace{-1truecm}

\begin{figure}[h]
\begin{center}
\psfig{file=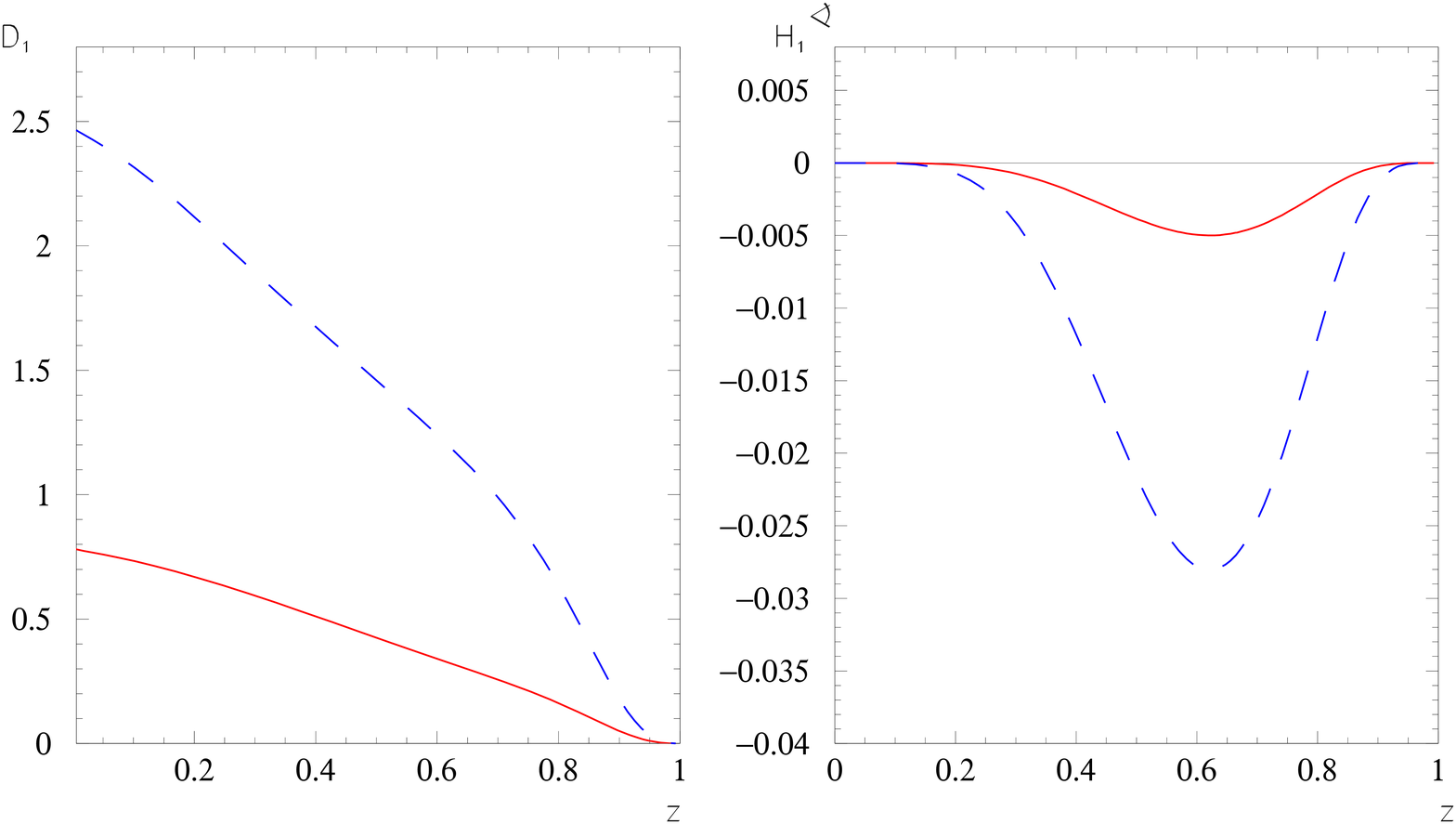, height=7cm, width=14cm}
\end{center}
\caption{\label{fig:ffplot} The FF $D_1 (z)$ (left) and 
  $H_{1\, (R)}^{\open} (z)$ (right). Solid line for $N_{q\rho} = 0.9$ GeV$^3$,
  and the integral~\protect{(\ref{eq:sumrule})} amounting to 0.14; 
  dashed line for $N_{q\rho} = 1.6$ GeV$^3$, and the integral 
  equals 0.48.}
\end{figure}

\section{Numerical Results}
\label{sec:results}

In the remainder of the paper, we present numerical results in the context 
of the spectator model for both the process-independent FF and the SSA of 
Eqs.~(\ref{eq:ssa-x}) and (\ref{eq:ssa-mh}) for semi-inclusive lepton-nucleon 
DIS. Considering different possible scenarios for $h_1$, we discuss 
the implications for an experimental search for transversity.

The input parameters of the calculation can basically be grouped in 
three classes:
\begin{itemize}
\item values of masses and coupling constants taken from phenomenology, as
  $m_\pi = 0.139$ GeV, $m_\rho = 0.785$ GeV, with $f_{\rho \pi \pi}$ and
  $\Gamma_\rho$ as described in Sec.~\ref{sec:vertices} and
  Sec.~\ref{sec:propagators}, respectively; 
  
\item values consistent with other works on the spectator model and the
  constituent quark model, as $\Lambda_\pi = 0.4$ GeV, $\Lambda_\rho = 0.5$
  GeV and $m_q = 0.34$ GeV~\cite{Jakob:1997wg,Bianconi:2000uc};
  
\item 
parameters, such as $N_{q\pi}$ and $N_{q\rho}$, without constraints 
that are firmly established, or at least usually adopted, in the
literature.
\end{itemize}

As previously anticipated in Sec.~\ref{sec:vertices}, the last ones are
constrained using the integral~(\ref{eq:sumrule}) and the
proportionality~(\ref{eq:qpi-coup}) derived from the Goldberger-Treiman
relation. All results will be plotted according to two extreme scenarios,
where the integral~(\ref{eq:sumrule})
amounts to 0.14 ($N_{q\rho} = 0.9$ GeV$^3$, corresponding
to solid lines in the figures) and 0.48 ($N_{q\rho} = 1.6$ GeV$^3$,
corresponding to dashed lines in the figures). Because of the high degree of
arbitrariness due to the lack of any data, the results should be interpreted
as the indication not only of the sensitivity of the considered observables to
the input parameters, but also of the degree of uncertainty that can be
reached within the spectator model. In the same spirit, when dealing with the
SSA of Eqs.~(\ref{eq:ssa-x},\ref{eq:ssa-mh}), $f_1$ and $h_1$ are calculated
consistently within the spectator model~\cite{Jakob:1997wg} or, 
alternatively, $f_1$ and $g_1$ are
taken from consistent parametrizations and $h_1$ is calculated again according
to two extreme scenarios: the nonrelativistic prediction $h_1 = g_1$ or the
saturation of the Soffer inequality, $h_1 = (f_1+g_1)/2$. The parametrizations
for $f_1, g_1,$ are extracted at the same lowest possible scale 
($Q^2 = 0.8$ GeV$^2$), consistently with the valence quark approximation 
assumed for the calculation of the FF. 

In Fig.~\ref{fig:ffplot} the integrated $D_1^u (z)$ and 
$H_{1\, (R)}^{\open \, u} (z)$ are shown. Again, we recall that the solid line
corresponds to a weaker $q\rho q$ coupling than the dashed line. The choice of
the form factors at the vertices also guarantees the regular behaviour at the
end points $z=0,1$. The strongest asymmetry in the fragmentation 
(recall that $H_1^{\open}$ is defined as the probability difference for the
fragmentation to proceed from a quark with opposite transverse polarizations)
is reasonably reached at $z\sim 0.4$. Once again, we stress that this result,
particularly its persistent negative sign, does not depend on a specific hard
process and can influence the corresponding azimuthal asymmetry. 

\begin{figure}[h]
\begin{center}
\psfig{file=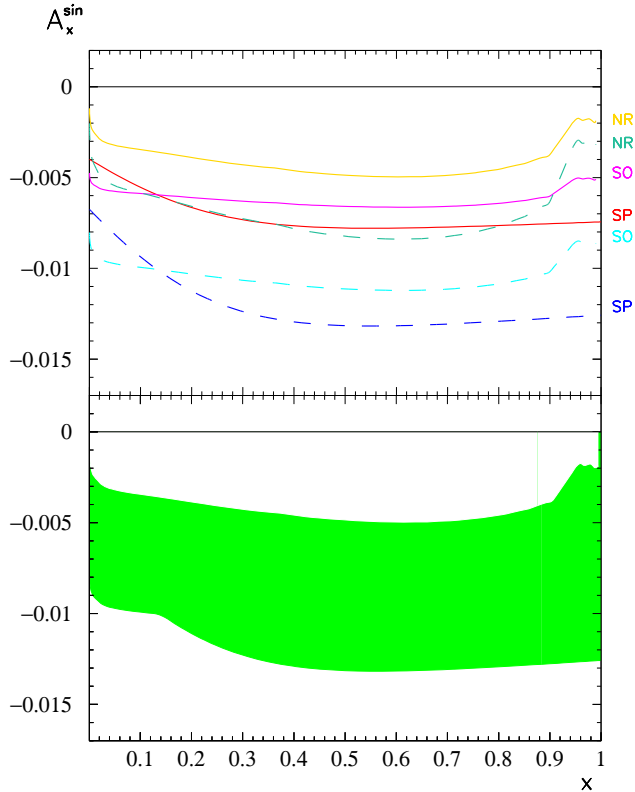, width=8.5cm}
\end{center}
\caption[dummy]{
  \label{fig:ssa-xplot} The SSA of Eq.~\protect{(\ref{eq:ssa-x})}. In
  the upper plot, solid lines and dashed lines as in
  Fig.~\protect{\ref{fig:ffplot}}. Label SP stands for DF calculated in the
  spectator model~\protect{\cite{Jakob:1997wg}}. Labels NR and SO indicate 
  $f_1$ from Ref.~\protect{\cite{Gluck:1998xa}} and $g_1$ from
  Ref.~\protect{\cite{Gluck:1996yr}}, but with $h_1 = g_1$ and 
  $h_1 = (f_1+g_1)/2$, respectively. In the lower plot, the corresponding 
  uncertainty band is shown (see text).
} 
\end{figure}

\begin{figure}[h]
\begin{center}
\psfig{file=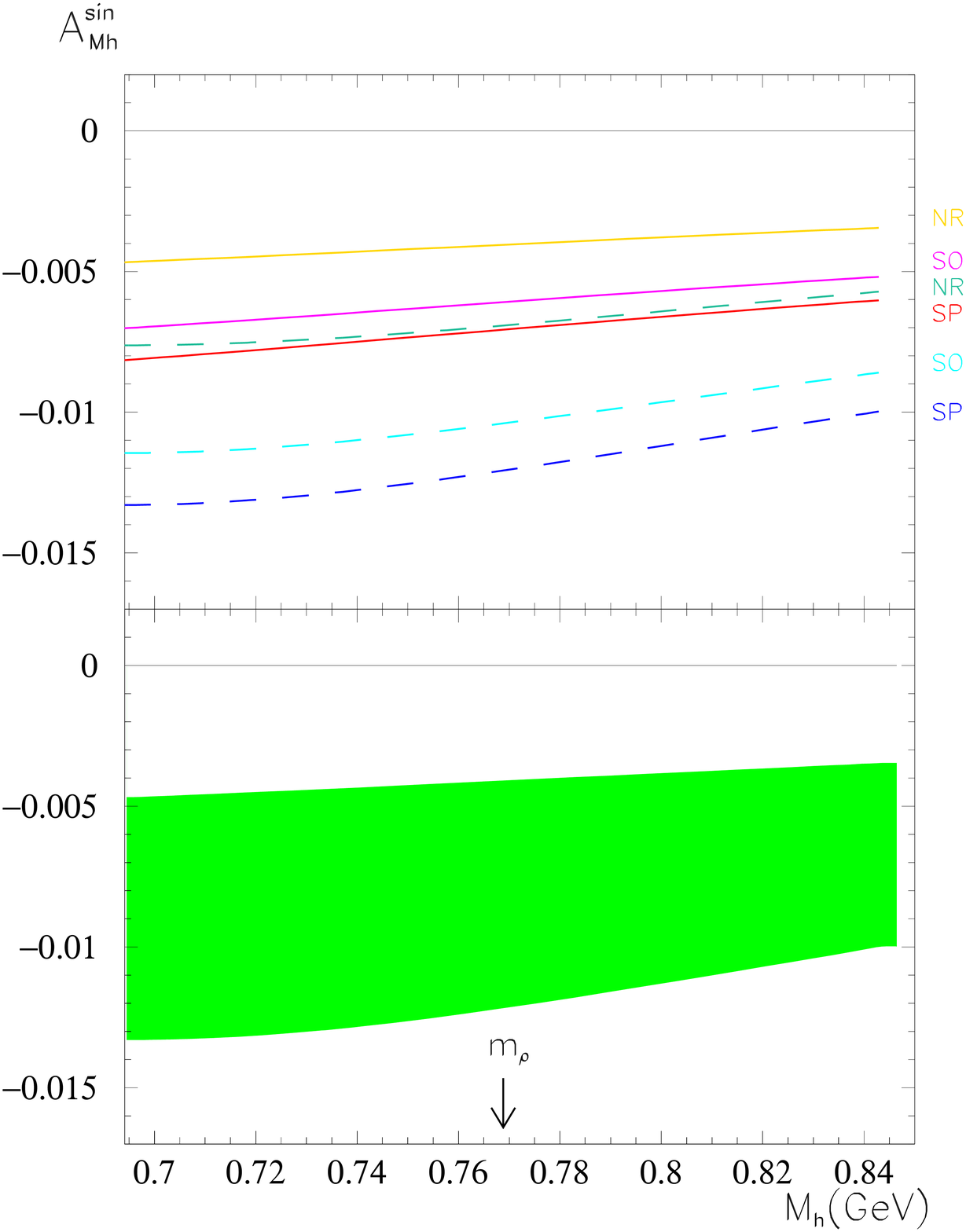, width=9cm}
\end{center}
\caption{\label{fig:ssa-mhplot} The SSA of Eq.~\protect{(\ref{eq:ssa-mh})}. In
  the upper plot, solid lines and dashed lines as in
  Fig.~\protect{\ref{fig:ffplot}}. Labels and meaning of lower plot as in
  Fig.~\protect{\ref{fig:ssa-xplot}}.} 
\end{figure}

In fact, the SSA~(\ref{eq:ssa-x}) and (\ref{eq:ssa-mh}) for two-pion 
inclusive lepton-nucleon DIS as shown in Figs.~\ref{fig:ssa-xplot} and 
\ref{fig:ssa-mhplot}, respectively, turn out to be negative due to the sign 
of $H_1^{\open\,u}$. The solid and dashed lines again refer to the weaker or 
stronger $q\rho q$ couplings in the FF, respectively. For each parametrization, 
three different choices of DF are shown. The label SP refers to the DF 
calculated in the spectator model~\cite{Jakob:1997wg}. The label NR indicates 
that $f_1$ and $g_1$ are taken consistently from the leading-order parametrizations of 
Ref.~\cite{Gluck:1998xa} and Ref.~\cite{Gluck:1996yr}, respectively, with 
$h_1 = g_1$. The label SO indicates the same parametrizations but with the 
Soffer inequality saturated, i.e.~$h_1 = (f_1 + g_1)/2$. 
In the lower plot of each figure the ``uncertainty band'' is shown as a
guiding line. It is built by taking, for each $z$ or $M_h$, the maximum and
the minimum among the six curves displayed in the corresponding upper
plot. The first obvious comment is that even the simple mechanism described in
Fig.~\ref{fig:diagspec} produces a measurable asymmetry. For the HERMES experiment 
the size of the asymmetry may be at the lower edge of possible measurements, 
given the observed rather small average multiplicity which does not favor the 
detection of two pions in the final state. On the other hand, the planned transversely 
polarized target clearly will improve the situation of azimuthal spin-asymmetry 
measurements compared to the present one. COMPASS or possible future experiments 
at the ELFE, TESLA-N, or EIC facilities will have less problems because of higher 
counting rates. The second important result is that the sensitivity of the SSA to the 
parameters of the model calculation for the FF and to the different parametrizations 
for the DF is weak enough that the unambigous message of a negative asymmetry 
emerges through all the range of both $x$ and  $m_\rho -\Gamma_\rho /2 = 0.69$ GeV
$\leq M_h \leq m_\rho +\Gamma_\rho /2 = 0.84$ GeV. In particular, we do not
find any change in sign for $A^{\sin \phi}_{M_h}$, contrary to what is
predicted in Ref.~\cite{Jaffe:1998hf}.  

\section{Outlooks}
\label{sec:end}

In this paper we have discussed a way for addressing the transversity
distribution $h_1$ that we consider most advantageous compared to other
strategies discussed in the literature. At present, the SSA seem
anyway preferable to the DSA. But the fragmentation of a transversely
polarized quark into two unpolarized leading hadrons in the same current jet
looks less complicated than the Collins effect, both experimentally and
theoretically. Collinear factorization implies an exact cancellation of the
soft divergencies, avoiding any dilution of the asymmetry because of Sudakov form 
factors, and in principle makes the QCD evolution simpler, though we have not 
addressed this subject in the present paper. The new effect,
that allows for the extraction of $h_1$ at leading twist through the new
interference FF $H_1^{\open}$, relates the transverse polarization of the
quark to the transverse component of the relative momentum of the hadron pair
via a new azimuthal angle. This is the only key quantity to be determined
experimentally, while the Collins effect requires the determination of the
complete transverse momentum vector of the detected hadron.  

We have shown also quantitative results for $H_1^{\open}$ in the case of
$\pi^+ \pi^-$ detection, and the related SSA for the example of
lepton-nucleon scattering, because modelling the interference
between different channels leading to the same final state is simpler than
describing the Collins effect, where a microscopic knowledge of the structure
of the residual jet is required. We have adopted a spectator model
approximation for $\pi^+ \pi^-$ with an invariant mass inside the $\rho$
resonance width, limiting the process to leading-twist mechanisms. The
interference between the decay of the $\rho$ and the direct production of
$\pi^+ \pi^-$ is enough to produce sizeable and measurable
asymmetries. Despite the theoretical uncertainty due to the arbitrariness in
fixing the input parameters of the calculation of FF and in choosing the
parametrizations for the DF, the unambigous result emerges that in the
explored ranges in $x$ and invariant mass $M_h$ the SSA are always negative
and almost flat. 

Anyway, it should be stressed again that, even if there are good arguments for
considering the mechanisms depicted in Fig.~\ref{fig:diagspec} a good
representation of $\pi^+ \pi^-$ production in the considered energy range,
still the calculation has been performed at leading twist and in a
valence-quark scenario. Therefore, higher-twist corrections and QCD evolution
need to be explored before any realistic comparison with experiments could be
attempted.

\acknowledgments

We acknowledge very fruitful discussions with Alessandro Bacchetta and Daniel 
Boer, in particular about the symmetry properties 
of the interference FF.\newline
This work has been supported by the TMR network HPRN-CT-2000-00130.

\end{document}